\documentclass{aa}  

\usepackage{graphicx}
\usepackage{txfonts}
\usepackage{subfigure}
\usepackage{subfig}
\usepackage{amssymb}
\usepackage{natbib}
\bibpunct{(}{)}{;}{a}{}{,}  
\usepackage{url}
\usepackage{color}

\begin{document} 

\titlerunning{Upper limits to interstellar NH$^+$ and  para-NH$_2^-$ abundances}
\title{Upper limits to interstellar NH$^+$ and  para-NH$_2^-$  abundances}
\subtitle{\emph{Herschel}\thanks{Herschel is an ESA space observatory with science instruments provided by European-led Principal Investigator consortia 
and with important participation from NASA}-HIFI  observations towards Sgr\,B2\,(M)  and  G10.6$-0.4$ (W31C)}  
\authorrunning{C.M.~Persson et~al.}    
  \author{C.M.~Persson     
          \inst{1},
M.~Hajigholi\inst{1},
 G.E.~Hassel\inst{2},
A.O.H.~Olofsson\inst{1}, 
          J.H.~Black\inst{1},
  E.~Herbst\inst{3},  
    H.S.P.~M\"uller\inst{4}, 
J.~Cernicharo\inst{5},  
E.S.~Wirstr\"om\inst{1},
M.~Olberg\inst{1}, 
\AA.~Hjalmarson\inst{1},   
D.C.~Lis\inst{6}, 
H.M.~Cuppen\inst{7}, 
 M.~Gerin\inst{8},    
K.M.~Menten\inst{9}         
          }   
   \offprints{carina.persson@chalmers.se}
   \institute{Chalmers University of Technology, Department of Earth and Space Sciences, Onsala Space Observatory,  SE-439 92 Onsala, Sweden.
    \email{\url{carina.persson@chalmers.se}} %1 Carina, Mitra, Eva, Henrik och JHB och Åke
\and Department of Physics \& Astronomy, Siena College, Loudonville, NY  12211,   USA  % 2 Hassel
\and   Department of Chemistry, University of Virginia, McCormick Road, Charlottesville, VA 22904, USA % 3  Herbst
\and I. Physikalisches Institut, Universit\"at zu K\"oln, Z\"ulpicher Str. 77, 50937 K\"oln, Germany% 4 Holger 
\and Centro de Astrobiolog\`{i}a, CSIC-INTA, 28850, Madrid, Spain % 5  Tom, Cernicharo, Javier  
\and California Institute of Technology, Cahill Center for Astronomy and Astrophysics 301-17, Pasadena, CA 91125, USA  % 6 Darek Lis
\and Radboud University Nijmegen, IMM - Faculty of Science, P.O. Box 9010, 6500 GL Nijmegen, The Netherlands  % 7 Herma Cuppen
\and LERMA-LRA, UMR 8112 du CNRS, Observatoire de Paris, \'Ecole Normale
Sup\'erieure, UPMC \& UCP, 24 rue Lhomond, 75231 Paris Cedex 05, France % 9  massimo och maryvonne
\and  Max-Planck-Institut f\"ur Radioastronomie, Auf dem H\"ugel 69, D-53121 Bonn, Germany  % 9 KM Menten & Wyrowski
}

   \date{Received March 4, 2014 /  Accepted May 21, 2014}

  \abstract
{The understanding of interstellar nitrogen chemistry has improved significantly with recent 
results from the \emph{Herschel} Space Observatory. To set even  better  constraints,  we   report  here
on deep searches for the  NH$^+$ ground state rotational transition \mbox{$J = 1.5 - 0.5$} of the  
$^2 \Pi _{1/2}$ lower spin ladder,  
with fine-structure transitions at 1\,013 and 1\,019~GHz, and the 
para-NH$^-_2$ \mbox{$1_{1,1} - 0_{0,0}$} rotational transition at 934~GHz 
  towards  Sgr\,B2\,(M) and  G10.6$-0.4$ (W31C) using
the \emph{Herschel} Heterodyne Instrument for the Far-Infrared (HIFI).    
No clear detections of NH$^+$   are  made and the derived upper limits  relative to the  
total number of hydrogen nuclei  are  
$\lesssim2\times$10$^{-12}$ and  $\lesssim7\times$10$^{-13}$   in the  
Sgr\,B2\,(M) molecular envelope   and in the G10.6$-0.4$   molecular cloud, 
respectively.  
The searches are, however, complicated by the fact that
 the 1\,013~GHz transition lies only
$-$2.5~km~s$^{-1}$ from a CH$_2$NH line, which is seen in absorption in Sgr\,B2\,(M),  
and that the  hyperfine structure components in the 1\,019~GHz transition are spread over 
134~km~s$^{-1}$.  
Searches for the so far undetected NH$^-_2$ anion turned out to be unfruitful towards G10.6$-0.4$, while 
the para-NH$^-_2$ $1_{1,1} - 0_{0,0}$ transition was tentatively detected towards Sgr\,B2\,(M) 
at a velocity of 19~km~s$^{-1}$. 
Assuming that the absorption occurs at the nominal source velocity of +64~km~s$^{-1}$, the rest frequency would be 933.996~GHz, offset by 141~MHz from our estimated value. 
Using this feature as an upper limit, we  found  \mbox{$N$(p-NH$^-_2$)$\lesssim4\times$10$^{11}$~cm$^{-2}$},  
which implies an abundance  of  $\lesssim8\times$10$^{-13}$ in  the Sgr\,B2\,(M) molecular envelope. 
The upper limits for both species in the  diffuse line-of-sight gas    
are less than 0.1 to 2~\% of the values found for NH,  NH$_2$, and NH$_3$ 
towards both  sources, and  the 
abundance  limits are      
 $\lesssim2-4\times$10$^{-11}$. 
An updated pseudo time-dependent chemical model with
constant physical conditions, including both gas-phase and surface
chemistry, 
predicts an NH$^+$ abundance a few times lower than our present upper limits 
in diffuse gas and under typical Sgr\,B2\,(M) 
envelope conditions.  
The NH$_2^-$ abundance is predicted to be several orders of magnitudes lower than our observed
limits, hence not supporting our tentative detection. 
Thus, while NH$_2^-$ may be very difficult to detect in interstellar space, 
it could, on the other hand,  
 be possible to 
detect NH$^+$ in  regions where   the ionisation rates  
of  H$_2$ and N  are greatly enhanced.  
}
 
   \maketitle
%
%________________________________________________________________

\section{Introduction}
 
An important species in the nitrogen chemistry, NH$^+$, has for a long time been awaiting   its first discovery.  
Besides 
its key chemical role in the reaction chain leading to more complex nitrogen-bearing species, 
 NH$^+$ has also been identified  as a potential 
candidate for probing variations in 
the fine-structure constant, $\alpha$, and electron-to-proton mass ratio, $\mu$ 
\citep{2011PhRvA..83f2514B}. 
Another undetected but interesting species in the nitrogen chemistry is the 
anion   NH$_2^-$.

Searches for NH$^+$ and  NH$_2^-$ are, however,   difficult not only because of their expected very low abundances, 
but also  since their strongest transitions lie at frequencies that are generally inaccessible to
ground-based telescopes. 
With the launch of \emph{Herschel} \citep{Pilbratt2010, 2012A&A...537A..17R} and its sensitive 
Heterodyne Instrument for the Far-Infrared (HIFI), which was designed to 
perform high-resolution  observations at frequencies    480-1250 and 1410-1910~GHz, 
searches for  the  fundamental rotational transitions of  NH$^+$ and NH$_2^-$ became possible.

Previous searches for NH$^+$ using \emph{Herschel}-HIFI  in the diffuse line-of-sight gas towards 
the high-mass star-forming regions G10.6$-0.4$ and W49N,  
resulted in average upper limits      of the NH$^+$ abundance relative to molecular hydrogen  \mbox{$\lesssim4\times$10$^{-10}$}, and \mbox{$N$(NH$^+$)/$N\mathrm{(NH)} \lesssim 4-7$~\%}
\citep{2012A&A...543A.145P}.

 \begin{table}[\!htb] 
\centering
\caption{Observed NH$^+$   and  para-NH$_2^-$   transitions. 
}
\begin{tabular} {lccccc  } 
 \hline\hline
     \noalign{\smallskip}
Species  	& Frequency\tablefootmark{a}	    
&   	  \multicolumn{2}{c} {$T_\mathrm{C}$\tablefootmark{b}}   
 &\multicolumn{2}{c}{$1\sigma$/$T_\mathrm{C}$\tablefootmark{c}}  
\\    \noalign{\smallskip}
&&    G10.6  &  SgrB2  &  G10.6  &  SgrB2    \\
  &(GHz) &            (K)& (K)& (K)& (K)      \\
     \noalign{\smallskip}
     \hline
\noalign{\smallskip}  

NH$^+$	  	&1\,012.540\tablefootmark{d}   	   	& 3.3	&	7.3 & 0.0019 & 0.0015	 \\
	 	&1\,019.211\tablefootmark{e}    	& 3.3  	&	7.4 & 0.0022 & 0.0040   \\
 \noalign{\smallskip}
p-NH$_2^-$ 	& 933.855\tablefootmark{f} 	  	&  2.6  & 	7.5 & 0.0035 & 0.0013\\ 

    \noalign{\smallskip} \noalign{\smallskip}
\hline 
\label{Table: transitions}
\end{tabular}
\tablefoot{
{ 
NH$^+$ was observed in band 4a with the 1\,013~GHz line in the lower sideband, and the
1\,019~GHz line in the upper sideband.
} 
\tablefoottext{a}{Frequencies without  the nuclear hyperfine structure \citep[hfs;][]{2009JChPh.131c4311H} 
that are used to convert frequencies to   
Doppler velocities relative to the local standard at rest 
$V_\mathrm{LSR}$. 
Hfs   
components of NH$^+$ can be found in  Tables~\ref{Table: 1013 hfs transitions}--\ref{Table: 1019 hfs transitions}. 
The \mbox{p-NH$_2^-$} frequency is estimated  using spectroscopic data from \citet{1986JChPh..85.4222T}. 
}
\tablefoottext{b}{The single sideband    continuum intensity.}  
\tablefoottext{c}{The rms noise (for a channel width of 1~km~s$^{-1}$)  divided by $T_\mathrm{C}$.} 
\tablefoottext{d}{$J = 1.5^- - 0.5^+$ ($-$ and +  denotes the parity).}
\tablefoottext{e}{$J = 1.5^+ - 0.5^-$.}
\tablefoottext{f}{$J_{K_a, K_c} = 1_{1,1} - 0_{0,0}$.}
}
\end{table}

 In this paper  we present the results of new,  deeper   searches for NH$^+$,  and for the first time also   for  NH$_2^-$,  
towards G10.6$-$0.4 and  Sgr\,B2\,(M). Both sources are very 
well-known star-forming regions and  extremely luminous sub-millimetre and infrared 
continuum sources. 
The  ultra-compact H~{\sc II} region
G10.6$-$0.4 in the star-forming  complex  W31 is   located in the Galactic   30~km~s$^{-1}$ arm at a 
 distance of 4.95~kpc \citep{2014ApJ...781..108S}, and  
the Sgr\,B2\,(M) region is one of the chemically rich sources close to the Galactic centre  at a distance of 8.5~kpc 
  \citep[e.g.][]{2000ApJS..128..213N, 2010A&A...521L..20B}. 
We also model the abundances of NH$^+$ and NH$_2^-$ under four different interstellar conditions  
with a pseudo time-dependent 
chemical model, and explore how the surface chemistry, cosmic ionisation rate, and  
assumed initial metal abundances influence the derived   abundances.

 \section{Spectroscopy, observations, and data reduction}\label{Section: observations and data reduction}
The observed transitions are listed in Table~\ref{Table: transitions}. 
Measurements  of the two fine-structure transitions   in the lowest \mbox{$N = 1 - 1, J = 1.5 - 0.5$} rotational  transition of NH$^+$ in its 
$^2 \Pi _{1/2}$ lower spin ladder 
were performed by  \citet{1986CPL...132..213V} and included 
resolved hyperfine structure (hfs). 
An energy-level diagram can be found in \citet[][their Fig.~1]{2009JChPh.131c4311H}. 
The Einstein $A$ values 
(\mbox{Tables~\ref{Table: 1013 hfs transitions}$-$\ref{Table: 1019 hfs transitions}})  
were derived by one of us  (HSPM) 
from these data, using the experimental ground-state electric dipole 
moment of 1.988~(28)~D \citep{2012PhRvA..85a2519M}, and taking   
additional parameters from \citet{2009JChPh.131c4311H} into account. 
The frequency of  the $^1A_1$, $J_{K_a, K_c} = 1_{1,1} - 0_{0,0}$ NH$_2 ^-$ 
transition  was taken from the  Madrid molecular spectroscopy excitation (MADEX) database 
\citep[][Table~\ref{Table: NH2m hfs transitions}]{2011.ECLA.cernicharo}. 
It was calculated from the spectroscopic parameters reported by \citet{1986JChPh..85.4222T}. 
The uncertainties of the  infrared transition frequencies are of the order of 100~MHz. 
While statistics may improve the prediction of transition frequencies, correlation 
among the spectroscopic parameters or vibration-rotation interaction may lead to 
significantly increased uncertainties. \citet{1993JChPh..99.8349B} calculated a ground-state 
dipole moment of 1.311~D with an estimated uncertainty of 0.01~D.

Emission or absorption features of other species were assigned by consulting the 
 Cologne Database for Molecular Spectroscopy
(CDMS) 
\citep{2001A&A...370L..49M, 2005JMoSt.742..215M}, 
Jet Propulsion Laboratory 
(JPL) \citep{1998JQSRT..60..883P}, or MADEX catalogues. Specifically, 
the NH$_2$ \citep{2001JMoSt.599..293G}, CH$_2$NH \citep{2012A&A...544A..19D}, and SO$_2$ \citep{2005JMoSp.232..213M}
entries were taken from the CDMS catalogue while the methanol entry \citep{2008JMoSp.251..305X} 
was taken from the JPL catalogue.

The observations, which took place in   April and September  2012, are summarised
 in  Table~\ref{Table: transitions}  and the  
observational identifications are found in the on-line Table~\ref{Table: obsid}.  
 We used the
dual beam switch mode and the wideband spectrometer (WBS) with a bandwidth of 
4$\times$1~GHz and an effective spectral  
resolution of 1.1~MHz ($\Delta v = 0.3$~km~s$^{-1}$). 
Two
orthogonal polarisations   
were used in all the observations. 
All lines towards G10.6$-$0.4 were observed with three different overlapping frequency settings of the local 
oscillator (LO)   
to
determine the sideband origin
of the lines since HIFI uses double sideband (DSB) receivers. 
Towards Sgr\,B2\,(M) we   used the spectral scan mode and eight different overlapping LO settings because of  
its extreme density of emission lines. 
 
 The pointings were centred at   $\alpha$\,=\,17$^\mathrm{h}$47$^\mathrm{m}$20\,$\fs$6, 
$\delta = -28^\circ$\,23$\arcmin$\,03.2$\arcsec$ ($J$2000) for  Sgr\,B2\,(M), and 
 $\alpha$\,=\,18$^\mathrm{h}$10$^\mathrm{m}$28\,$\fs$7, 
$\delta$\,=\,$-$19$^\circ$\,55$\arcmin$\,50.0$\arcsec$ ($J$2000) 
for G10.6$-$0.4. The source systemic velocities are  58--69  and $-3$~km~s$^{-1}$   
 for 
Sgr\,B2\,(M)  and G10.6$-$0.4, respectively. Absorptions in the source molecular 
clouds are centred at 
+64 and -0.5~km~s$^{-1}$, and the foreground gas along the respective sight-line 
is detected in absorption   from $-$140 to 27, and 10 to 55~km~s$^{-1}$.

The data were reprocessed using the  \textit{hifiPipeline} task in   HIPE   
 version 9.0,   up to  level 2 providing fully calibrated  DSB  spectra  for   G10.6$-$0.4  on   
the $T_\mathrm{A}^*$  
 antenna temperature   
intensity scale where the lines are calibrated to single sideband (SSB) and the continuum to DSB. 
For the  Sgr\,B2\,(M)  observations,  
we used in addition    the \textit{doDeconvolution} task up to level 2.5    to   provide 
fully calibrated  SSB  spectra. 
The   \emph{FitHifiFringe} task    
was then used 
to fit and remove residual ripples in the spectra, except for the NH$^+$ 1\,019~GHz data towards Sgr\,B2\,(M) since this spectrum had too many spectral features.
The G10.6$-$0.4 data quality is excellent with very low intensity ripples,  
with  good agreement between the
three LO-tunings, and without any visible contamination from the image sidebands.

The FITS files were exported to  the spectral line   
software package   
{\tt xs}\footnote{{\tt http://www.chalmers.se/rss/oso-en/observations/data-\\reduction-software}},  
which was used in the subsequent data reduction. 
All  tunings and both polarisations were included in the averaged noise-weighted spectra for all transitions,  
which were convolved to a channel width of 1~kms$^{-1}$.
Baselines of order five were removed from the G10.6$-$0.4 spectra, and of order 
three and seven from the 
Sgr\,B2\,(M) NH$_2^-$ and NH$^+$ 1\,013~GHz spectra (average $T_\mathrm{C}$ added afterwards). 
No baseline was removed from the  Sgr\,B2\,(M) NH$^+$ 1\,019~GHz spectrum.

 \section{Results} \label{section: results}
 
\begin{figure} 
\centering
\resizebox{\hsize}{!}{ 
\includegraphics{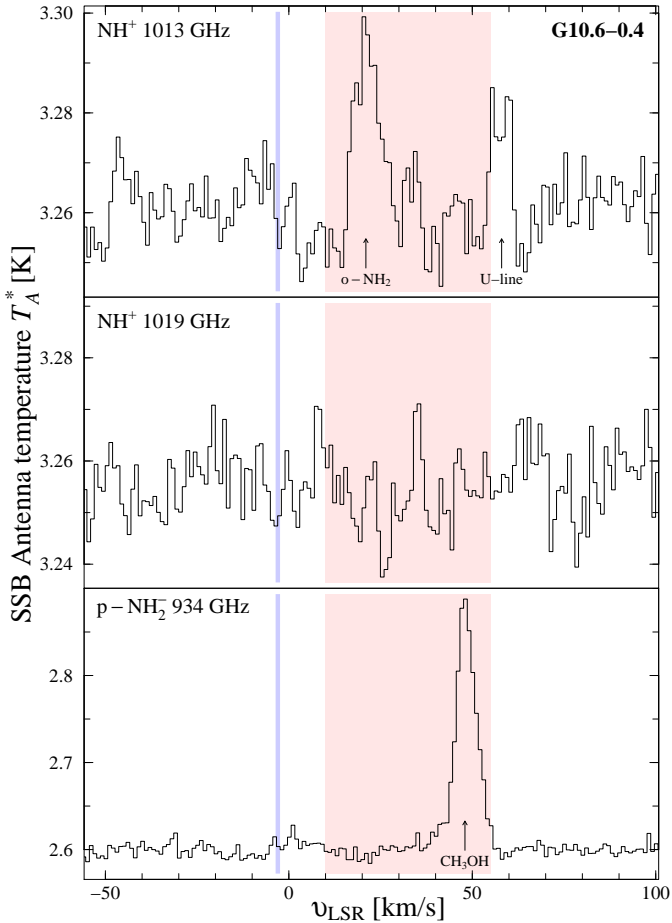}}  
\caption{G10.6$-$0.4:  searches for NH$^+$   and 
p-NH$_2^-$   show  no detections at the source velocity $-$3~km~s$^{-1}$ (marked in blue) 
 or the line-of-sight absorption  at 10--55~km~s$^{-1}$ (marked in red). 
The ortho-NH$_2$ and CH$_3$OH emission lines originate in the G10.6$-$0.4 molecular cloud.   
}
\label{Fig: W31C all species}
\end{figure}

\begin{figure} 
\centering
\resizebox{\hsize}{!}{ 
\includegraphics{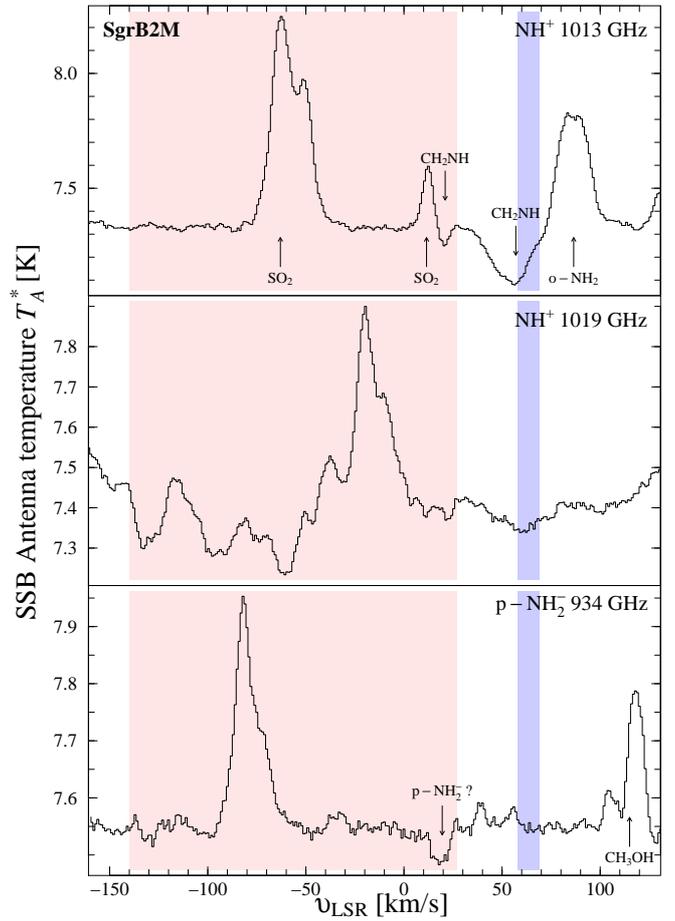} 
}
\caption{Sgr\,B2\,(M):  searches for NH$^+$   and 
p-NH$_2^-$   show  no clear detections at the source molecular cloud velocities 58--69~km~s$^{-1}$ (marked in blue) 
 or the line-of-sight absorption   between $-$140 and 27~km~s$^{-1}$ (marked in red). 
Possible weak NH$^+$ absorption from the molecular cloud is blended with the stronger
CH$_2$NH absorption. 
Para-NH$_2^-$ is tentatively detected from the source molecular cloud, however, at 
\mbox{$V_\mathrm{LSR} =19$~km~s$^{-1}$}.  
All emission lines originate in the Sgr\,B2\,(M) molecular cloud. 
}
\label{Fig: SgrB2M all species}
\end{figure}

Figures~\ref{Fig: W31C all species}--\ref{Fig: SgrB2M all species}   show  the    
averaged WBS spectra of all observed transitions as a function of 
  the local standard of rest velocity, $V_\mathrm{LSR}$. 
The continuum  and rms are given in Table~\ref{Table: transitions}.
We   performed an unbiased  search  for emission  and absorption lines from  NH$^+$ and 
para-NH$_2^-$ in the source molecular clouds, and  absorption  from diffuse or translucent gas 
along the  lines-of-sight. 
Despite the low noise levels,   no detections are found in the G10.6$-$0.4 data  
(Fig.~\ref{Fig: W31C all species}).
Column densities  are therefore  
3$\sigma$ upper limits, estimated with a typical line width   (4~km~s$^{-1}$) and the total 
line-of-sight velocity range. 
The emission line visible in the NH$^+$ 1\,013~GHz band  (upper panel) at +21~km~s$^{-1}$ is identified as 
ortho-NH$_2$  \mbox{$4_{2,2}- 4_{1,3}$}.

Towards  Sgr\,B2\,(M), we find two  absorption features at   $V_\mathrm{LSR}\approx60$~km~s$^{-1}$ in both NH$^+$  spectra 
(upper and middle panels in Fig.~\ref{Fig: SgrB2M all species}).  
It is to be noted that the
1\,013 and 1\,019~GHz lines are  
expected to show    very different
line profiles since they both have 14 hfs components  and are 
spread over velocity ranges of 26 and 134~km~s$^{-1}$, respectively.  
To check whether  the observed 
line profiles fit the   respective NH$^+$ hfs,     
we 
model   the absorption  of both  lines using  Gaussian  optical depth profiles
generated for each  hfs component. These profiles are  made to fit the observations 
under the condition that 
the $V_\mathrm{LSR}$ and line width are the same for both transitions 
\citep[cf. Method~I in][]{2012A&A...543A.145P}.   
Assuming a  sideband gain ratio of unity, we calculate the 
line opacities  as 
\mbox{$\tau\!=\!-\ln{(T_\mathrm{A}^*/T_\mathrm{C}})$},  
where $T_\mathrm{A}^*$ is the SSB antenna temperature. 
As seen in Fig.~\ref{Fig: SgrB2M Gauss models NHp 1013 1019} (on-line material) the fit 
to the 1\,013~GHz line shows  
a very good agreement with the observed line profile, whereas the  1\,019~GHz fit suggests 
either that   the detection is not real, or that a 
considerable part of the absorption  comes from
other species, or that it is (partly)  caused by remaining ripples 
which we were not able to remove.

\begin{table*}[\!ht] 
\centering
\caption{Resulting  NH$^+$ and NH$_2^-$ column densities, $N$, column density ratios with  related species, and abundances, $X$, with respect to the total amount of hydrogen 
    towards Sgr\,B2\,(M) 
and G10.6$-$0.6.
}
\begin{tabular} {lcccccccccc  } 
 \hline\hline
     \noalign{\smallskip}
& \multicolumn{10}{c}{Line-of-sight\tablefootmark{a}}  \\  \noalign{\smallskip}
Source  & $V_\mathrm{LSR}$ & $N(\mathrm{NH}^{+})$ & {\normalsize$\frac{N(\mathrm{NH}^{+})}{N(\mathrm{NH})}$} & {\normalsize$\frac{N(\mathrm{NH}^{+})}{N(\mathrm{NH_2})}$}\tablefootmark{b}  &
 {\normalsize$\frac{N(\mathrm{NH}^{+})}{N(\mathrm{NH_3})}$}\tablefootmark{c} &  $N$(p-NH$_2^-)$  & {\normalsize$\frac{N(\mathrm{p-NH_2^-})}{N(\mathrm{NH_2})}$}\tablefootmark{b}  &
 {\normalsize$\frac{N(\mathrm{p-NH_2^-})}{N(\mathrm{NH_3})}$}\tablefootmark{c}  & 
$X(\mathrm{NH}^{+})$ & $X$(p-NH$_2^-)$\\
 \noalign{\smallskip}
  & (km~s$^{-1}$)	& (cm$^{-2}$) &(\%) &(\%) &(\%)  &  (cm$^{-2}$) &(\%)  &(\%)  \\
     \noalign{\smallskip}
     \hline
     \noalign{\smallskip}

Sgr\,B2  &  $-$140-27  	& $\lesssim1.7\times$10$^{12}$&  $\lesssim0.1$&  $\lesssim0.2$ &	$\lesssim0.2$& $\lesssim 9.9\times$10$^{11}$  & 
$\lesssim0.1$ &$\lesssim0.1$ & $\lesssim4\times$10$^{-11}$\tablefootmark{d} & 	 $\lesssim2\times$10$^{-11}$\tablefootmark{d}	 \\ 
G10.6  & 10-55  &$\lesssim1.6\times$10$^{12}$	& $\lesssim0.9$  &  $\lesssim1.5$  
 	& $\lesssim1.9$  &   $\lesssim2.2\times$10$^{12}$ &$\lesssim1.9$& $\lesssim2.6$& $\lesssim3\times$10$^{-11}$\tablefootmark{e}    & $\lesssim4\times$10$^{-11}$\tablefootmark{e}\\ 
  \noalign{\smallskip} \noalign{\smallskip}\hline
  \noalign{\smallskip}   \noalign{\smallskip}

& \multicolumn{10}{c}{Source molecular cloud/envelope}   \\  \noalign{\smallskip}\noalign{\smallskip}

Sgr\,B2  &  57--68  	& $\lesssim9.1\times$10$^{11}$& $\ldots$ &   $\ldots$ &	 $\ldots$& $\lesssim4.0\times$10$^{11}$  &  $\ldots$  &   $\ldots$ & $\lesssim2\times$10$^{-12}$\tablefootmark{f} & 	 $\lesssim8\times$10$^{-13}$\tablefootmark{f}	 \\ 

G10.6  & $-$3  &$\lesssim2.8\times$10$^{12}$ 	&  $\ldots$  &   $\ldots$ 
 	&  $\ldots$ &   $\lesssim2.0\times$10$^{12}$ &  $\ldots$  &   $\ldots$& $\lesssim7\times$10$^{-13}$\tablefootmark{g}    & $\lesssim5\times$10$^{-13}$\tablefootmark{g}\\ 
   \noalign{\smallskip}

\hline 
\label{Table: columns and abundances}
\end{tabular}
\tablefoot{ 
\tablefoottext{a}{The limits are estimated over  the velocity ranges  listed in column 2.}
\tablefoottext{b}{Using the high temperature  ortho-to-para (OPR) limit of 3 for NH$_2$.} 
\tablefoottext{c}{Using OPR(NH$_3) = 0.7$   \citep{2012A&A...543A.145P}.}
\tablefoottext{d}{\mbox{$N_\mathrm{H}= 2N(\mathrm{H_2}) + N(\mathrm{H})  = 4.3 \times 10 ^{22}$~cm$^{-2}$} \citep{1989ApJ...338..841G, 2011ApJ...734L..23M}.}
\tablefoottext{e}{\mbox{$N_\mathrm{H}=   2N(\mathrm{H_2}) + N(\mathrm{H}) = 5.6\times10 ^{22}$~cm$^{-2}$}  
($N(\mathrm{H})$ from  Winkel et~al. (in prep.) and $N(\mathrm{H_2})$ from \citet[][]{2012A&A...540A..87G})}. 
\tablefoottext{f}{$N_\mathrm{H}\approx 2\times N(\mathrm{H}_2)  = 5\times$10$^{23}$~cm$^{-2}$ in the Sgr\,B2\,(M)
molecular \emph{envelope} \citep{1989ApJ...337..704L}.}
\tablefoottext{g}{$N_\mathrm{H}= 2\times N(\mathrm{H}_2)/2 = 4\times$10$^{24}$~cm$^{-2}$  in the  
G10.6$-$0.4  molecular cloud 
  \citep{1978ApJ...221L..77F}   assuming that NH$^+$ and NH$_2^-$  are most likely to be seen in absorption    thereby 
probing half of the total column density that is in front of the continuum source.}  
} 
\end{table*}

Despite
the agreement of the fits, we ascribe the largest part of  the feature in the 1\,013~GHz spectrum  to  
CH$_2$NH. This species has two  transitions with similar line strengths close to   NH$^+$ at 1\,013~GHz:      
$3_{3,1}-2_{2,0}$ 
(with $E_l= 41$~K)  only +8.5~MHz ($-$2.5~km~s$^{-1}$) from the  
NH$^+$ line, and $3_{3,0}-2_{2,1}$ 
seen as a  narrow absorption   at 
21~km~s$^{-1}$ in Fig.~\ref{Fig: SgrB2M all species} (upper panel). This line, however,   blends with the 
SO$_2$  
$41_{5,37}-40_{4,36}$ (1\,012.673~GHz) emission and is therefore   easily missed.  
Our   identification is also supported by previous observations of  CH$_2$NH  in both absorption 
($1_{1,1}-0_{0,0}$) and emission towards Sgr\,B2\,(M)   \citep{2000ApJS..128..213N}. 
We    modelled the SO$_2$ emission,    
both CH$_2$NH absorptions,   
and  the o-NH$_2$ \mbox{$4_{2,2}- 4_{1,3}$} 
emission line wing (seen at $\sim$70~km~s$^{-1}$ in Fig.~\ref{Fig: SgrB2M all species}, upper panel)
in order to subtract these lines in the 
search for any remaining weak NH$^+$ absorption. 
More details of the modelling are found in on-line Sect.~\ref{on-line section: ch2nh and so2 modelling}, and  
all modelled lines are shown in Fig.~\ref{Fig: SgrB2M all lines gauss models}.  After subtraction of
the modelled lines, we find a weak remaining absorption feature  at   $V_\mathrm{LSR}$\,=\,69~km~s$^{-1}$ with  an integrated opacity of 0.08~km~s$^{-1}$. 
This feature  is  considered  as an upper limit of 
NH$^+$  in the SgrB2\,(M) molecular envelope. An unidentified  remaining absorption feature is also seen at 45.5~km~s$^{-1}$ with an integrated opacity of 0.19~km~s$^{-1}$. 
Both these features are, however, very weak and may   well be remaining artefacts from our modelling or ripples 
in the baseline.

In the 934 GHz band
we find an unidentified absorption feature
at  $V_\mathrm{LSR}\approx +18.5$~km~s$^{-1}$ towards Sgr\,B2\,(M) with a line width 
of $\approx\!9$~km~s$^{-1}$ and an integrated opacity of  0.09~km~s$^{-1}$
(lower panel in Fig.~\ref{Fig: SgrB2M all species}). 
This feature is used as an upper limit to the  para-NH$_ 2^-$ % tentatively assigned to 
$1_{1,1} - 0_{0,0}$ line. 
If the absorption is caused by para-NH$_2^-$, it implies a rest frequency of 
933.973--934.009~GHz for this transition, which is
 118--154~MHz higher than our estimated  frequency  assuming that the  nominal 
source velocity is  between 56 and 68~km~s$^{-1}$. 
Results of quantum chemical calculations on  NH$_2 ^-$ were recently 
reported employing high level coupled cluster calculations with additional correction 
and large basis sets with extrapolation to infinite basis set size \citep{2009JChPh.131j4301H}. 
Using their best spectroscopic parameters, a frequency of 932.726~GHz is 
derived for the $J = 1_{1,1} - 0_{0,0}$ 
transition. The level of agreement with the value derived from the experimental spectroscopic 
parameters corresponds to the one expected under favourable conditions and does not permit  
exclusion of the Sgr\,B2\,(M) absorption feature as being potentially due to NH$_2 ^-$.

We convert the upper limits of NH$^+$ opacities to  
column densities with the non-equilibrium 
homogeneous radiative transfer code 
{\tt{RADEX}}\footnote{
{\tt http://www.sron.rug.nl/{$\sim$}vdtak/radex/index.shtml}}  \citep{2007A&A...468..627V}  
  to 
correct for possible population of molecules in unobserved excited levels. 
We use 
\mbox{$n(\mathrm{H_2}) = 70$~cm$^{-3}$} and a kinetic temperature \mbox{$T_\mathrm{K} = 100$~K} 
for the diffuse line-of-sight conditions, and 
\mbox{$n(\mathrm{H_2}) = 10^2 - 10^4$~cm$^{-3}$} and   \mbox{$T_\mathrm{K} = 20-40$~K} for the denser 
envelopes of the source molecular clouds. The results are not 
very sensitive to changes in density because of the high critical density of the nitrogen hydrides \mbox{($n_\mathrm{crit}\sim10^8$~cm$^{-3}$)}. 
For the line-of-sight,
we use the
average Galactic background radiation in the solar neighbourhood
plus the cosmic microwave background radiation as background radiation field.  
In addition, for the source molecular clouds we  include their 
respective observed spectral energy distribution.

Since  no  collisional coefficients are available for  NH$_ 2^-$,  we estimate the column density  of 
molecules in the ground-state   
using 
\begin{equation}
\resizebox{.9\hsize}{!}{$N_l(\mathrm{p-NH}_2^-) = 8\pi  \frac{\nu^3}{c^3}   \frac{g_l}{g_u\,A_{ul}}  \int \tau \mathrm{d}V =  4.7 \times 10^{12} \int \tau \mathrm{d}V \, \, \, \mathrm{[cm^{-2}]} \ .  $}    
\end{equation}

The resulting upper limits for the column density and abundance with  
 respect to the total column of hydrogen   
  are 
found in Table~\ref{Table: columns and abundances}. 
Here we also present limiting abundance ratios, relative to the chemically 
related species NH, NH$_2$, and NH$_3$ along the line-of-sight   gas towards both sources.  
Column densities towards G10.6$-$0.4 (averaged over \mbox{$V_\mathrm{LSR}=10-55$~km~s$^{-1}$})
are taken from  \citet{2012A&A...543A.145P}.  
A full spectral scan 
of Sgr\,B2\,(M) using \emph{Herschel}-HIFI 
has been performed by the HEXOS  Key Programme   \citep{2010A&A...521L..20B}.  From  these data, spectra of  
 the ground-state rotational transitions of NH, NH$_2$, and NH$_3$
were extracted and 
 compared to our data.  Resulting ratios are consistent with the findings in \citet{2010A&A...521L..45P}, and  $N(\mathrm{NH_3})$ is in agreement with \citet{2010A&A...522A..19W}.
We reduced  these  data  
in a manner similar to that  described 
in Sect.~\ref{Section: observations and data reduction} for our Sgr\,B2\,(M) data,  and    fitted Gaussian optical
depth profiles to the absorption 
lines to estimate    the column densities. % \citep[cf.][]{2012A&A...543A.145P}. %+60~km$^{-1}$ component.

\begin{figure}
\begin{center}
 \subfigure{
\includegraphics[scale=.33]{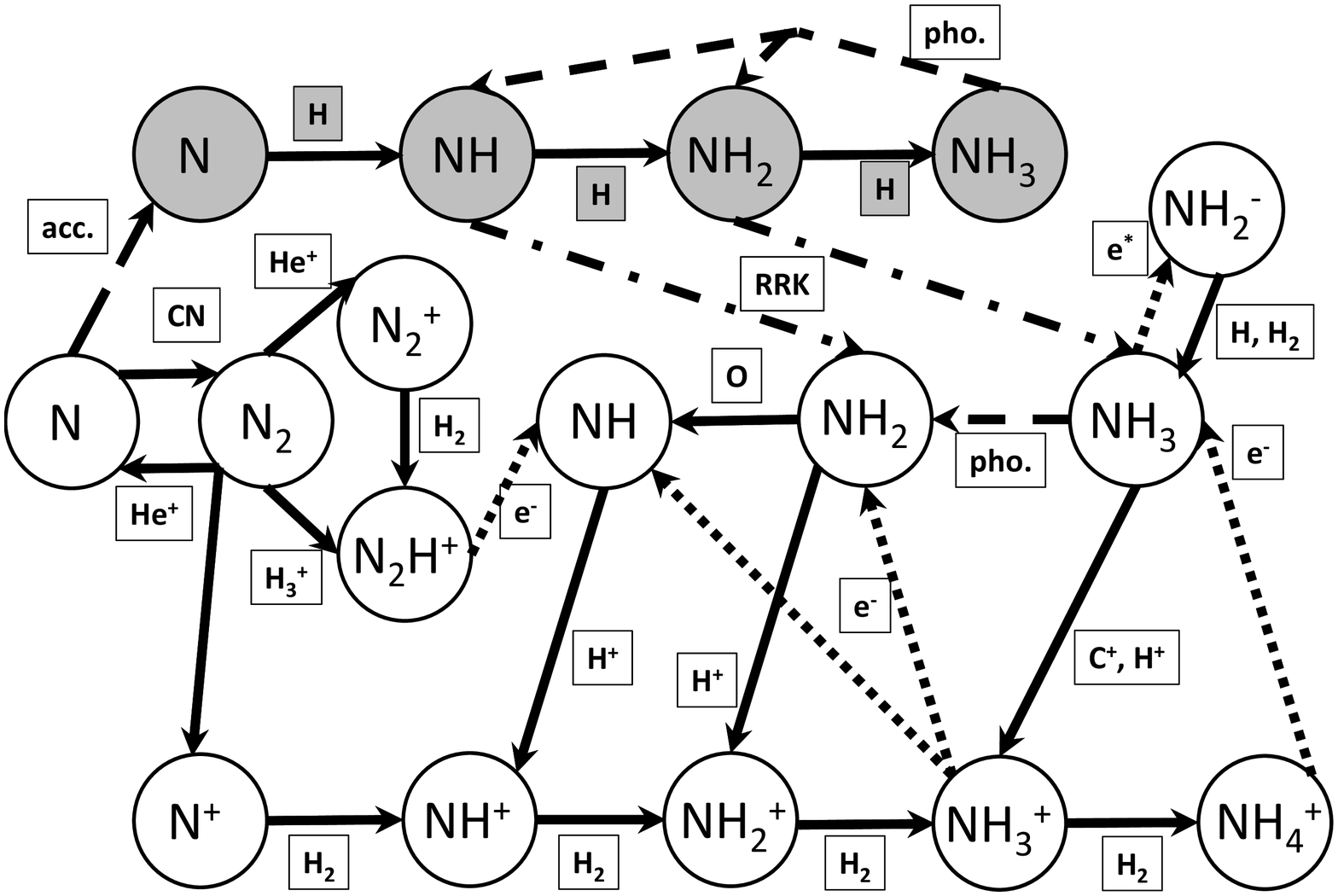}} %7166Rxns/NH.hiT.eps} }
\vspace{.3in}
\subfigure{
\includegraphics[scale=.33]{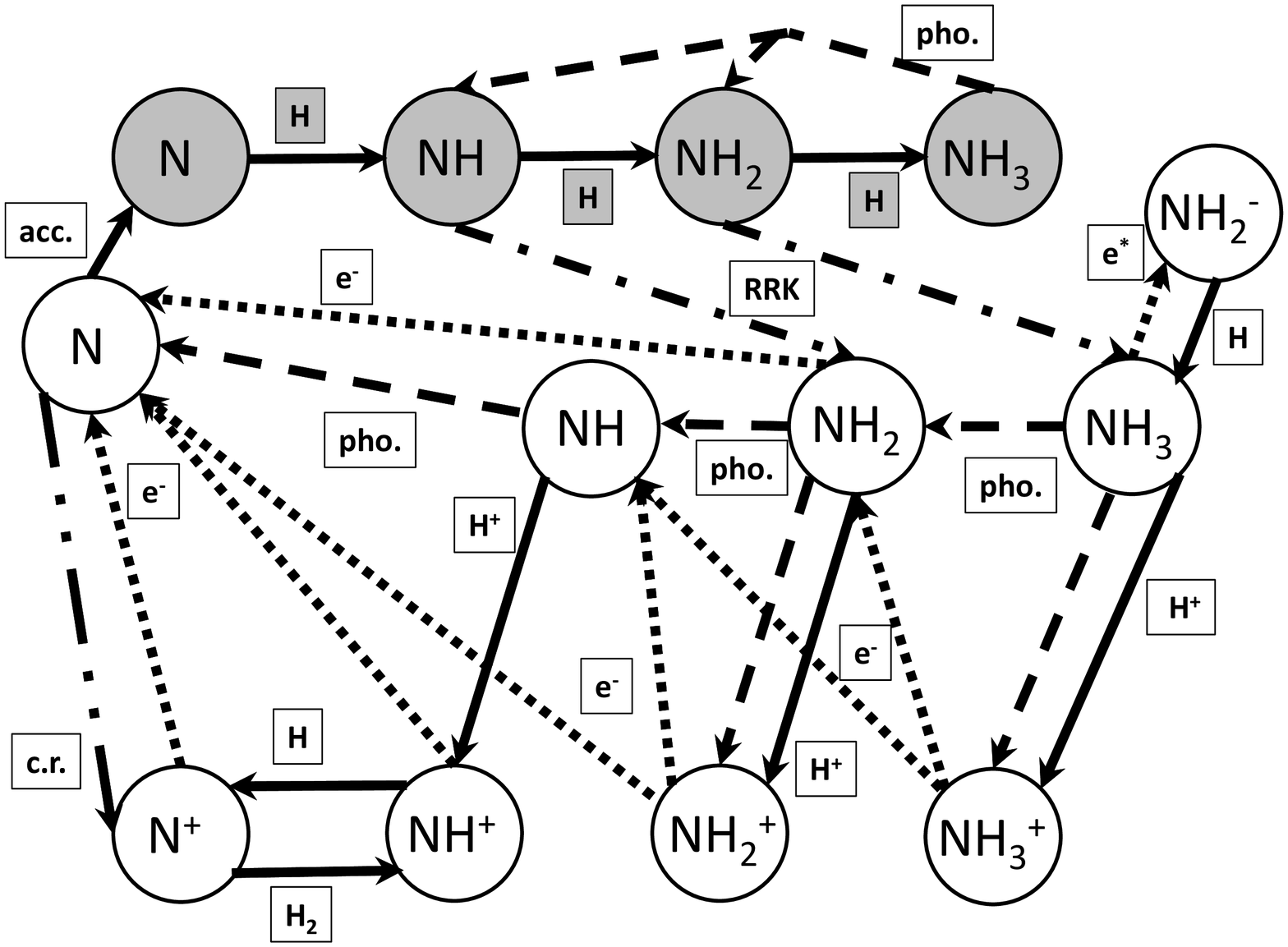}}  %7166Rxns/NH2.hiT.eps} }
\caption{\emph{Major} formation paths of  nitrogen hydrides   at
  $t=10^{6}$ years for   \emph{translucent}  gas conditions  (upper panel)  and \emph{diffuse} gas conditions 
(lower panel).    Gas phase species appear   in white and   
grain surface  species  in grey.  
Solid lines indicate gas phase reactions, dashed lines  photo-dissociation,  dotted lines
  dissociative recombination, electronic radiative recombination and attachment processes, 
and   dot-dashed lines   
  non-thermal desorption from the grain surfaces.      
\label{fig_chemnetwork}}
\end{center}
\end{figure}

\section{Chemical modelling} \label{section: discussion}

In \citet{2010A&A...521L..45P} we modelled  abundances of the nitrogen hydrides 
with a pseudo time-dependent chemical model with constant physical conditions, taking   both 
the gas-phase 
and grain surface chemistry into account,   using the Ohio State University (OSU) gas-grain
code \citep{1992ApJS...82..167H}. 
The     
predicted   NH$^+$ abundance was at most \mbox{$\sim\!10^{-13}-10^{-14}$} 
in translucent clouds with  $A_\mathrm{V}=2-3$.

In this paper, we have updated the chemical models 
to include the formation and destruction of NH$_2^-$.  
The expanded reaction network also includes high temperature reactions
 \citep{2010ApJ...721.1570H, 2011ApJ...743..182H} and presently considers 7\,176 reactions involving a total of 669
gaseous and surface species.  
 
Figure~\ref{fig_chemnetwork} shows two chemical networks outlining the \emph{major} reaction 
pathways involving nitrogen hydrides at $t=10^{6}$ years, under  translucent   and diffuse gas conditions. 
In both cases, the production of NH$^+$ purely by gas-phase processes largely depends on a sufficient source 
of N$^+$, which can be formed by cosmic ray ionisation of N or by reactions of He$^+$ with N$_2$ or CN. It 
should be noted  
that H$_3^+$ does not react rapidly with N, hence the latter route is the most important one in denser gas, 
while the former dominates in diffuse gas. In standard gas-phase ion-molecule chemistry, NH$^+$ 
then initiates 
the production of nitrogen hydrides (NH, NH$_2$, and NH$_3$) via subsequent reactions with H$_2$ and electron 
recombination.  This is, however, not effective in diffuse gas where hydrogen exists mostly in atomic form. Included 
in the chemical networks of Fig.~\ref{fig_chemnetwork} is also the surface reaction pathway for the formation 
of nitrogen hydrides, in which H atoms are 
added to N, and NH$_3$ is in turn   destroyed by photo-dissociation. 
The species NH$_2$ and NH$_3$  
can then be liberated  into  the gas phase through non-thermal desorption 
via the Rice-Ramsperger-Kassel (RRK) mechanism \citep{2006FaDi..133...51G, 2007A&A...467.1103G},  where the species desorb as
a result of exothermic surface reactions with an efficiency governed by the 
parameter $a_\mathrm{RRK}$, which is typically set to 0.01.
We have also considered the addition of photodesorption of NH$_3$ with an assumed yield 
of $Y_\mathrm{PD} = 10^{-3}$~molecules/UV~photon in the absence of a measured yield, based on the 
formulation of \citet{2007ApJ...662L..23O} for desorption of CO by both the direct interstellar 
radiation field and the field caused by cosmic rays.  The direct photodesorption process is 
of secondary importance to the RRK mechanism for the formation of NH$_3$(gas) in the 
translucent gas models and of even lesser importance in the diffuse gas models.
In the diffuse model, the major form of nitrogen is predominantly elemental N at  10$^6$~yrs, 
while in the the translucent model, elemental N, NH$_3$(ice), and N$_2$(gas) are the major forms.  In addition to the processes shown here, there are some minor processes, such as 
$\mathrm{O} + \mathrm{CN} \rightarrow \mathrm{CO} + \mathrm{N}$, 
$\mathrm{C} + \mathrm{NO} \rightarrow \mathrm{N} + \mathrm{CO}$, 
 and  NO$^+ + \mathrm{e}^- \rightarrow \mathrm{N} + \mathrm{O}$,      
that return some elemental N to the gas phase from less abundant forms, 
but these are omitted from Fig.~\ref{fig_chemnetwork}.

The abundance of NH$^+$ thus directly depends on a N$^+$ source and  the cosmic ionisation rate
 $\zeta(\mathrm{H}_2)$.  
Radiative recombination of N$^+$ is a slow process; therefore, in the presence of just a small H$_2$ 
fraction, N$^+$ is removed mainly by
\mbox{N$^+ + \mathrm{H}_2\!\rightarrow\,\mathrm{NH}^+ +\mathrm{H}$} , 
which is the source reaction of NH$^+$. Thus almost every cosmic ray ionisation of N will produce NH$^+$. 
Reactions with H$_2$ and with electrons removes NH$^+$, but the reactions with H$_2$ 
dominate as long as 
\mbox{e$^-$/H$_2\lesssim0.001$}. Within this limit,  %, we can estimate directly 
 independent of density and temperature, 
the NH$^+$ fractional abundance   is of the order of 10$^{-12}$ at $n(\mathrm{H}_2)\approx 100$~cm$^{-3}$
and $X(\mathrm{N})\approx 10^{-4}$.

The   NH$_2^-$  anion can  form   
via the dissociative attachment process through electron-impact on NH$_3$ 
 \begin{equation}\label{NH2m from NH3}
e^* + \mathrm{NH}_3 \rightarrow \mathrm{NH}_2^- + \mathrm{H} \ , 
\end{equation}
where $e^*$ represents an energetic electron. 
The energetic threshold for this process is  \mbox{$\varepsilon=3.857$~eV}.  
The possible destruction processes of NH$_2^-$ include photo-detachment, reactions with H$_2$, and 
mutual neutralisation in reactions with the most abundant
positive ions. 
If the reaction with H$_2$  is the dominant loss
process, then the density of  NH$_2^-$ at $T=50-100$~K will be of the order of
\begin{equation}
n(\mathrm{NH}_2^-) \sim 7\times 10^{-8}\,\frac{n(\mathrm{NH}_3)}{n(\mathrm{H}_2)}\ \ \mathrm{[cm^{-3}]}\ , 
\end{equation}
which immediately suggests a very low NH$_2^-$ abundance. 
A second formation route  of NH$_{2}^{-}$ is via slow radiative attachment of electrons to NH$_{2}$  
\begin{equation}  
{\rm NH_{2} + e^{-} \longrightarrow NH_{2}^{-}   + h\nu}\ . 
\end{equation}
  Finally, if the anion is formed in a local region rich in atomic 
rather than molecular hydrogen, it can be destroyed by associative attachment with atomic hydrogen, 
 \begin{equation}
  {\rm NH_{2}^{-} + H   \longrightarrow NH_{3} + e^{-},}
  \end{equation}
   or competitively via photo-detachment.  
Details about the different formation and destruction routes of NH$_{2}^{-}$ are found in
the on-line appendix~\ref{appendix: NH2m chemistry}.

 \begin{figure*}[htb]
\centering
\subfigure{
\includegraphics[angle=-90,scale=.35]{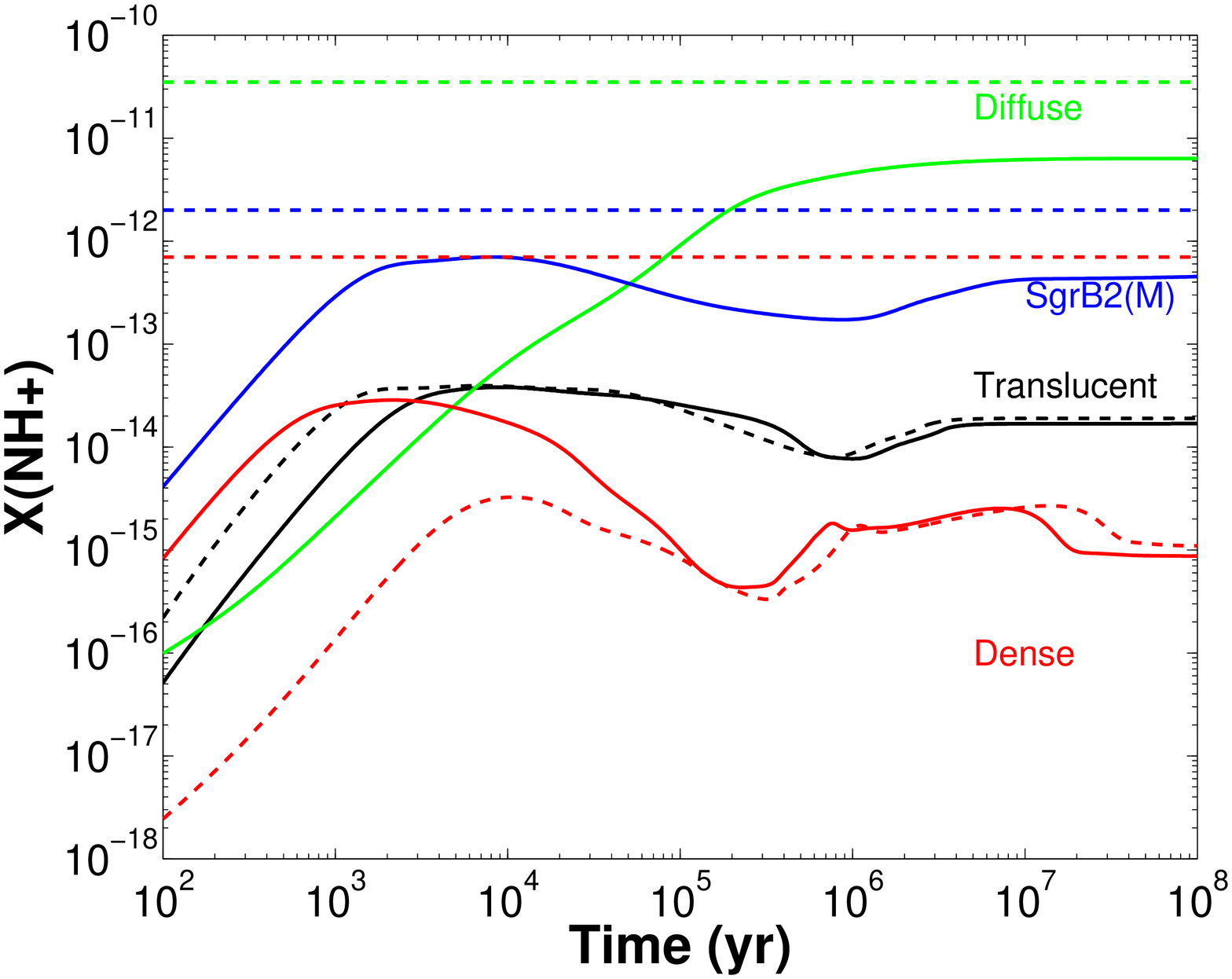}}
\subfigure{
\includegraphics[angle=-90,scale=.35]{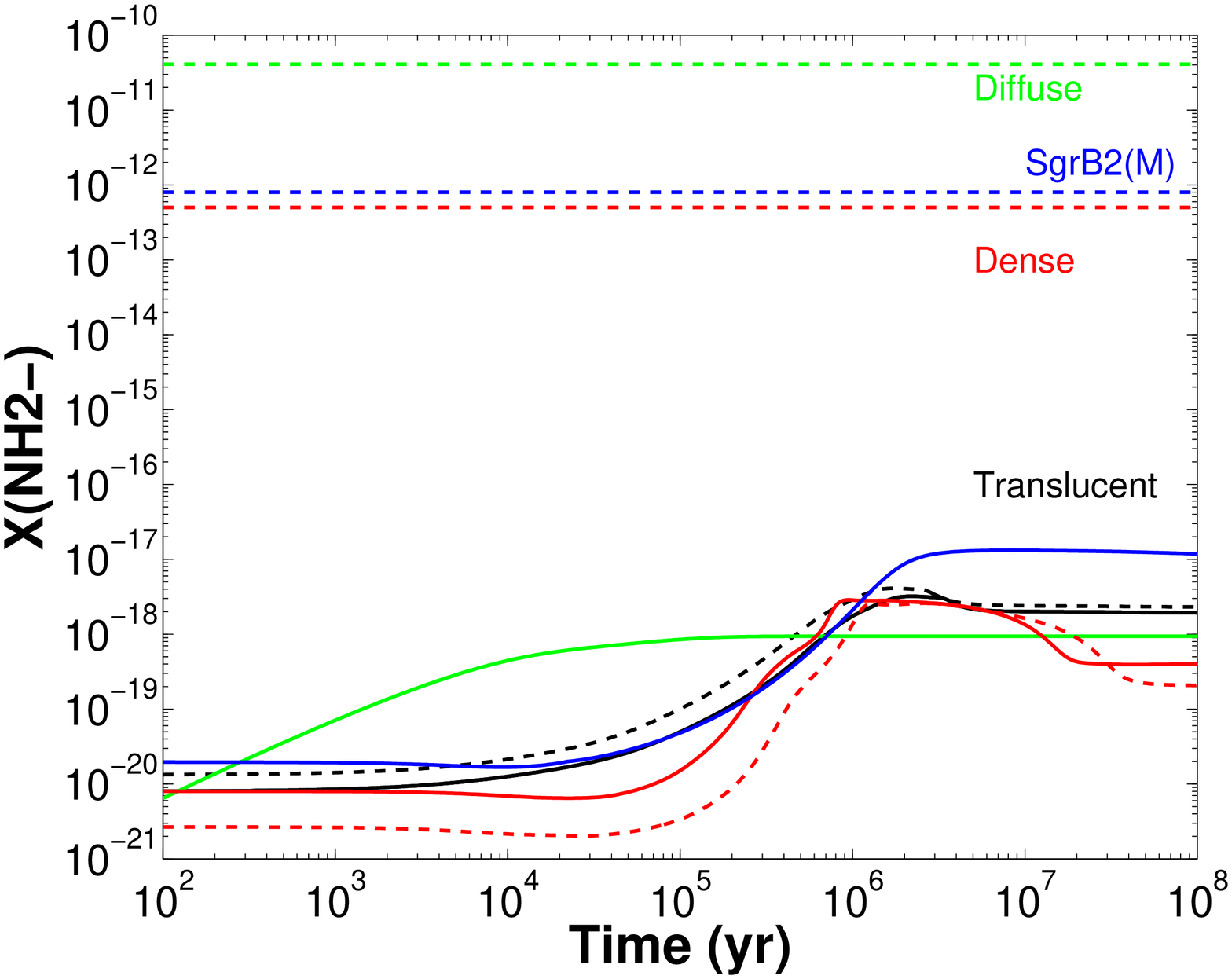}}
\vspace{.3in}
\caption{
Temporal evolution of the  NH$^+$ (left) and NH$_2^-$ (right) abundances using gas and surface chemistry    
and a typical active non-thermal desorption efficiency \mbox{$a_\mathrm{RRK} = 0.01$} using  four  different sets of 
conditions (see Table~\ref{Table: chemical model parameters}). 
Typical  \emph{diffuse} cloud conditions in green,  
\emph{translucent} cloud conditions in black, 
\emph{dense}  cloud conditions in red, and a special model for the 
\emph{Sgr\,B2\,(M) molecular envelope} in blue. 
The solid and dashed lines represent 
$T_\mathrm{K} = 30$~K and 50~K (translucent model) and $T_\mathrm{K} = 30$~K and 10~K (dense model), respectively. The observed upper limits from both species (Table~\ref{Table: columns and abundances}) 
in dense gas  (from W31C),  
diffuse gas (similar limits for both sight-lines), and in the Sgr\,B2\,(M) molecular
envelope 
 are indicated by dashed horizontal 
lines following the respective colour code.  
}
\label{Figure: 4}
\end{figure*}

 \begin{table*}[\!htb] 
\centering
\caption{Chemical models.
}
\begin{tabular} {lcccccc  } 
 \hline\hline
     \noalign{\smallskip}
Model 	&  $A_\mathrm{V} $ & $n_\mathrm{H}$\tablefootmark{a} &  $T_\mathrm{K}$&  $T_\mathrm{dust}$  & $\zeta(\mathrm{H}_2)$\tablefootmark{b} &  Metallicity\tablefootmark{c}\\
   \noalign{\smallskip}
& (mag)    &   (cm$^{-3}$)  &       (K)& (K)& (s$^{-1}$)       \\
     \noalign{\smallskip}
     \hline
\noalign{\smallskip}  

Diffuse\tablefootmark{d}  	&  0.5	&  70		&  100	&	17	& $2\times10^{-16}$ & High \\	
Translucent\tablefootmark{d}	&  3	& $5\times 10^3$& 30 \& 50	&	10	& $1.3\times10^{-17}$ &  Low \\	
Dense\tablefootmark{d}		&  10	&  $2\times 10^4$	& 10 \& 30 	&	10	& $1.3\times10^{-17}$
&  Low \\	
 Sgr\,B2\,(M) envelope\tablefootmark{e}  & 300   &  4$\times 10^3$   &   40 &	 10 	&  4$\times10^{-16}$  &  High  \\	
    \noalign{\smallskip} \noalign{\smallskip}
\hline 
\label{Table: chemical model parameters}
\end{tabular}
\tablefoot{
{All models use a ultraviolet flux of 1~$G_\mathrm{0}$.} 
\tablefoottext{a}{The total hydrogen density $n_\mathrm{H}= 2\,n(\mathrm{H_2}) + n(\mathrm{H})$.} 
\tablefoottext{b}{The cosmic ionisation rate of molecular hydrogen  \citep[cf.][]{2012ApJ...758...83I}.} 
\tablefoottext{c}{The assumed initial metal abundances   are found 
in  Table~\ref{Initial elemental abundances}.}
\tablefoottext{d}{Typical conditions in respective cloud type \citep[e.g.][]{2006ARA&A..44..367S}.}
\tablefoottext{e}{\citet{1989ApJ...337..704L}, \citet{2006A&A...454L..99V} 
and    \citet{2014ApJ...785...55O}.}
}
\end{table*}

Figure~\ref{Figure: 4}  shows the resulting temporal evolution of the  NH$^+$ and NH$_2^-$ 
abundances using the updated chemical network        
and a typical active non-thermal desorption efficiency \mbox{$a_\mathrm{RRK} = 0.01$}.  
We have modelled both species under  four  different physical conditions:
 typical \emph{dense gas},  
 \emph{translucent gas}, \emph{diffuse gas}, 
and a special model for the \emph{Sgr\,B2\,(M) molecular envelope}. 
All model parameters are 
found in Table~\ref{Table: chemical model parameters}.  

The choice of initial elemental abundance values vary for the different models.  For the translucent and dense models, we adopt ``low metal'' initial abundances, whereas we 
adopt a set of ``high metal'' initial abundances for the diffuse and    Sgr\,B2\,(M) models.   
The ``low metal'' values account for the incorporation of elements into refractory grains 
on the basis of observations of $\zeta$ Oph \citep{1982ApJS...48..321G, 2007A&A...467.1103G}, 
and the ``high metal'' values were developed to estimate the abundances if all of this 
material initially existed in the gas phase \citep{2008ApJ...680..371W, 2006A&A...457..927G}.  
The different values are listed in  Table~\ref{Initial elemental abundances}, and the 
effects of the adoption on abundance of NH$^+$ is explored in Fig.~\ref{onlineFigB7}, 
where   the choice of high metal abundances can be seen to  increase the abundance 
of NH$^+$ towards the observational upper limit for the diffuse model. 
The same trend is found for the Sgr\,B2\,(M) and dense models, however,   not for the   translucent model.

The dense models are   representative of the massive  
sources themselves, the translucent cloud conditions are traced by, for example, the NH, NH$_2$, and 
NH$_3$ absorptions along the sight-lines \citep{2012A&A...543A.145P}, and the diffuse cloud conditions 
are representative of  the line-of-sight 
clouds, from where we believe NH$^+$  originates. 
The Sgr\,B2\,(M) model reflects the very special conditions found in this source.  
In the on-line  Fig.~\ref{onlineFigB4} we show the  four  models again, but this time with the  
addition of NH, NH$_2$, and NH$_3$ for comparative purposes. 
We note that the above results for the translucent model are very similar to the models from  
\citet{2010A&A...521L..45P}.

To check how the surface chemistry affects the resulting abundances we also show 
\emph{(i)} gas and surface chemistry and 
inactive non-thermal desorption  efficiency $a_\mathrm{RRK} = 0$;  
\emph{(ii)} gas and surface chemistry
 with two different $a_\mathrm{RRK}$ desorption efficiencies;  and \emph{(iii)}
pure gas phase chemistry for     the translucent model in Fig.~\ref{onlineFigB5} (on-line material). 
This figure, as well as Fig.~\ref{fig_chemnetwork},  illustrates  that 
the surface chemistry with 
reactive desorption is a key factor in the formation of NH$_2$  and NH$_3$, followed in importance by 
the addition of electrons to NH$_4^+$. 
We find, however,  
that our models are not very sensitive to the exact value of the desorption probability, since the  model  
 with  $a_\mathrm{RRK} = 0.1$  gives very similar 
results to the model using $a_\mathrm{RRK} = 0.01$.

Since a high cosmic ray ionisation rate is crucial for the production of NH$^+$,  
our results are sensitive to its assumed value. In the on-line Figure~\ref{onlineFigB6} 
we     show,  therefore, 
how the  NH$^+$ abundance varies in the diffuse model using seven different values of  $\zeta(\mathrm{H_2})$. 
  The  
 NH$^+$ abundance increases by more than an order of magnitude 
and reaches the observational limit 
 when  $\zeta(\mathrm{H_2})$ 
increases from $10^{-17}$ to 10$^{-14}$~s$^{-1}$. 
 The other models also show similar trends.

The rotational transitions of NH$^+$ observed by us provide a sensitive way to search 
for this ion in absorption. There are also other kinds of transitions 
suitable, however, for interstellar absorption studies as pointed out by \citet{1982A&A...113..199D}. 
For example, the ground-state $\Lambda$-doubling transition at 13.6~GHz is observable 
from Earth and now has well determined hfs frequencies \citep{2009JChPh.131c4311H}. If 
we neglect hfs in both the lowest pure-rotational and $\Lambda$-doubling transitions 
and consider a column density of $N({\rm NH}^+)\sim10^{12}$ cm$^{-2}$, then the integrated 
optical depths in the 1\,013 and 1\,019~GHz lines are $\int \tau dv \sim 0.15$~km~s$^{-1}$ 
while the corresponding integrated optical depth at 13.6~GHz is 
$\sim0.005$~km~s$^{-1}$.  The electronic transitions at blue and ultraviolet wavelengths 
are slightly more sensitive than the 13.6 GHz transition. Based on the oscillator 
strengths tabulated by \citet{1982A&A...113..199D}, the corresponding values of 
$\int \tau dv$ are 0.009, 0.01, and 0.007~km~s$^{-1}$ in the 
A~$^2\Sigma^-$ - X $^2\Pi_{\rm r}$ (0,0), (1,0), and (2,0) bands near 464, 434, and 
410~nm wavelength, respectively. The ultraviolet band C~$^2\Sigma^+$ - X $^2\Pi_{\rm r}$ 
(0,0) near 289~nm will yield $\int \tau dv \sim 0.009$~km~s$^{-1}$ under the same 
conditions. The existing upper limits on visible and UV lines of NH$^+$ 
\citep{1976ApJ...204L.127S, 1980IAUS...87..247S, 1973ApJ...181L.122J} 
were not very sensitive compared to 
the submm-wave  results presented here. As far as we are aware, no limits have been 
derived from more modern optical data. Possible archival spectra of stars 
behind diffuse molecular clouds in which equivalent widths of $W_{\lambda} \lesssim 1$~m\AA\ 
could easily be measured in the 434~nm band  corresponding to a column 
density $\lesssim 6\times 10^{11}$~cm$^{-2}$, slightly better than the best limit 
in Table~\ref{Table: columns and abundances}.

\section{Conclusions}
Our derived NH$^+$ upper limits 
are an order of magnitude lower    
than previous estimates
\citep{2012A&A...543A.145P}. On the other hand,   
our chemical modelling suggests that the NH$^+$ abundance may 
still be a few times   lower than our present limits  in diffuse gas and under typical 
Sgr\,B2\,(M) molecular envelope conditions, and several orders of magnitude  lower   in 
 translucent and dense gas. 
Since a high ionisation rate is crucial for high NH$^+$ abundances,  
future searches should focus on regions with greatly enhanced ionisation rates 
\citep[cf.][]{2012ApJ...758...83I}. 
Searches for NH$^+$ are, however, complicated by the fact that one of  its lowest rotational 
transitions at 1\,013~GHz lies only $-$2.5~km~s$^{-1}$ from the   
$3_{3,1} - 2_{2,0}$ CH$_2$NH line seen in absorption in Sgr\,B2\,(M).

 In contrast to NH$^+$, the   NH$_2^-$ anion has  very low abundances in all models, 
not supporting our tentative detection in Sgr\,B2\,(M). This suggests  
that this species will be very difficult to detect in interstellar space.

\begin{acknowledgements}
HIFI has been designed and built by a consortium of institutes and university departments from across Europe, Canada, and the United States under the leadership of SRON, Netherlands Institute for Space Research, Groningen, The Netherlands and with major contributions from Germany, France, and the US. Consortium members are: Canada: CSA, U.Waterloo; France: CESR, LAB, LERMA, IRAM; Germany: KOSMA, MPIfR, MPS; Ireland, NUI Maynooth; Italy: ASI, IFSI-INAF, Osservatorio Astrofisico di Arcetri-INAF; Netherlands: SRON, TUD; Poland: CAMK, CBK; Spain: Observatorio Astronómico Nacional (IGN), Centro de Astrobiología (CSIC-INTA). Sweden: Chalmers University of Technology - MC2, RSS \& GARD; Onsala Space Observatory; Swedish National Space Board, Stockholm University - Stockholm Observatory; Switzerland: ETH Zurich, FHNW; USA: Caltech, JPL, NHSC.
C.P., J.H.B., and E.S.W. acknowledge generous support from the Swedish National Space Board. 
E.H. acknowledges the support of NASA for research related to the Herschel HIFI programme. 
H.S.P.M. is very grateful to the Bundesministerium f\"ur Bildung und
Forschung (BMBF) for financial support aimed at maintaining the
Cologne Database for Molecular Spectroscopy, CDMS. 
H.M.C acknowledges the European Research Council (ERC-2010-StG, Grant Agreement no. 259510-KISMOL) for financial support.
Support for this work was provided by NASA through an award issued by JPL/Caltech. 

\end{acknowledgements}

\bibliographystyle{aa-package/bibtex/aa}
\bibliography{references}

\Online
\appendix
 
\section{Modelling the CH$_2$NH absorption lines} \label{on-line section: ch2nh and so2 modelling}

 \begin{figure} 
\resizebox{\hsize}{!}{ 
\includegraphics{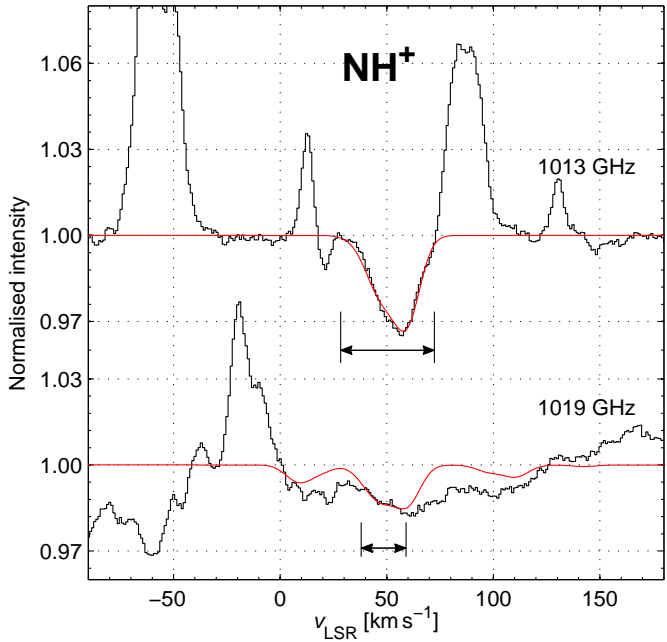}   
}
\caption{Sgr\,B2\,(M). Gaussian fits of NH$^+$ 1\,013 and 1\,019~GHz line profiles, including all hfs components,   to the observations. The arrows mark the velocity range in which the fit was made. 
}
\label{Fig: SgrB2M Gauss models NHp 1013 1019}
\end{figure} 

 \begin{figure} 
\resizebox{\hsize}{!}{ 
\includegraphics{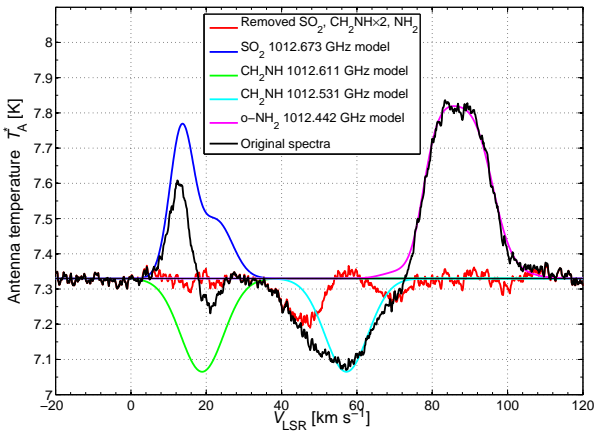}    
}
\caption{Sgr\,B2\,(M). 
Observed  
NH$^+$ 1\,013~GHz search spectrum (black) along with models of the two
absorbing CH$_2$NH transitions (green and cyan),  the expected SO$_2$ (blue) 
and NH$_2$ (pink) emissions, and the remaining spectrum 
after removal of their expected contributions (red; more details in Sect.~\ref{section: results}).
}
\label{Fig: SgrB2M all lines gauss models}
\end{figure}

CH$_2$NH has two  transitions with similar line strengths close to   NH$^+$ at 1\,013~GHz:      
$3_{3,1}-2_{2,0}$ (1012.531~GHz) with $E_l= 41$~K  only +8.5~MHz ($-$2.5~km~s$^{-1}$) from  
NH$^+$, and $3_{3,0}-2_{2,1}$ (1012.661~GHz)
seen as a  narrow absorption   at 
21~km~s$^{-1}$. It should be noted that this line     blends with the 
SO$_2$  
$41_{5,37}-40_{4,36}$ (1\,012.673~GHz) emission.  
We  model  the SO$_2$ emission,   
both CH$_2$NH absorptions,   
and  the o-NH$_2$ \mbox{$4_{2,2}- 4_{1,3}$} 
emission line wing (seen at $\sim$70~km~s$^{-1}$)  in order to subtract these lines in the 
search for any remaining weak NH$^+$ absorption. 
All modelled lines are shown  in Fig.~\ref{Fig: SgrB2M all lines gauss models} together with  
the original
data in black and the resulting spectra after subtraction of the above described lines is shown in red.  

The  numerous   SO$_2$ lines observed in our band 
(e.g. $39_{5,35}-38_{4,34}$ and $43_{5,39}-42_{4,38}$), as well as in the HEXOS spectral line survey,
 are used to reconstruct the SO$_2$ 1\,012.673~GHz line. The true CH$_2$NH absorption   is then found by 
  comparing the reconstructed  SO$_2$ emission with 
the observed line profile. 
Finally, 
the modelled CH$_2$NH absorption  is used as a template for the  CH$_2$NH absorption at 1012.531~GHz 
  since their line strengths are similar. 
We use {\tt RADEX} to check our modelled CH$_2$NH lines together with the 
225~GHz  ($1_{1,1}-0_{0,0}$)   line observed in absorption by \citet{2000ApJS..128..213N}. The integrated opacities 
of these three lines  
are matched using
a density of $n(\mathrm{H}_2) \sim 10^5$~cm$^{-3}$, a kinetic temperature  of  $\sim100$~K, a column density
$N(\mathrm{CH_2NH})\sim 1\times10^{15}$~cm$^{-2}$, and a line width of $\sim15$~km~s$^{-1}$ which supports
the above modelling and results.
After subtracting all modelled lines, we find a weak remaining absorption feature  at   $V_\mathrm{LSR}$\,=\,69~km~s$^{-1}$ with  an integrated opacity of 0.08~km~s$^{-1}$, which is used as an upper limit to NH$^+$.

 \begin{figure} 
\resizebox{\hsize}{!}{ 
\includegraphics{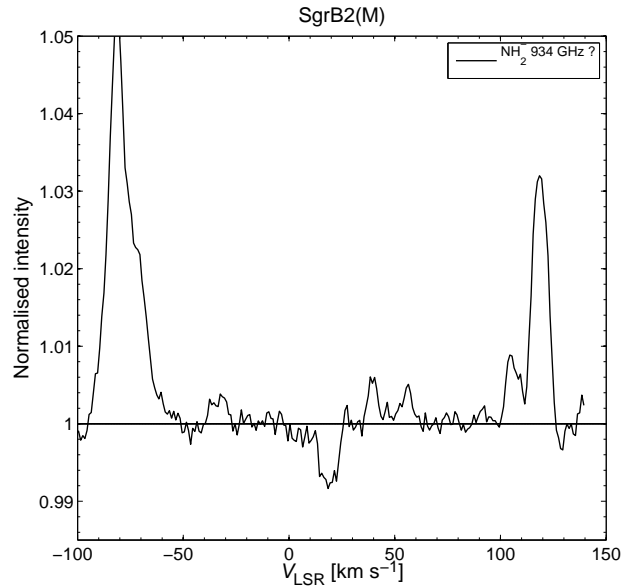}}  
\caption{The absorption feature found at a velocity of +18.5~km~s$^{-1}$ is tentatively  
identified as p-NH$_2^-$ from the source molecular cloud. We use the resulting integrated opacity as an upper limit to $N$(NH$_2^-$). 
The  
emission line at 110~km~s$^{-1}$ is methanol  9$_{2,7}$--8$_{1,8}$   (rest frequency 933.693~GHz).
}
\label{Fig: normalised nh2m}
\end{figure}

\section{NH$_2^-$ chemistry} \label{appendix: NH2m chemistry}

The   NH$_2^-$  anion can  form   
via the dissociative attachment process through electron-impact on NH$_3$ 
 \begin{equation}\label{NH2m from NH3}
e^* + \mathrm{NH}_3 \rightarrow \mathrm{NH}_2^- + \mathrm{H} \ , 
\end{equation}
where $e^*$ represents an energetic electron. 
The energetic threshold for this process is  \mbox{$\varepsilon=3.857$~eV}, potentially leading to an unusually high  production rate. 
At kinetic temperatures of
the order of 100~K or less, the thermal electrons have a characteristic energy
less than 9~meV. 
Therefore, the hot electrons required to form the anion are
extremely superthermal. A self-consistent treatment of the electron speed
distribution in the weakly ionised interstellar medium is currently being investigated
\citep[][in prep]{2014Black}. The crucial energy range for reaction (\ref{NH2m from NH3}) is 3.8
to 8.5~eV. In photon-dominated regions, including diffuse molecular
clouds, such electrons are produced mainly by the same photoelectric effect
involving dust and large molecules that dominates the heating of the gas. 
Energetic electrons are thermalised primarily by collisions with neutrals 
(H and H$_2$), rather than by elastic collisions
with thermal electrons, as long as the fractional ionisation is less than 10$^{-3}$.
The cross-section,  $\sigma_\mathrm{DA}$, for the dissociative attachment process   has a peak value of 
1.6~Mb near $\varepsilon=5.8$~eV, 
with vanishing values at $\varepsilon<4.2$ and at $\varepsilon>8.5$~eV \citep{1969JChPh..50.3024S, 2008IJMSp.277...96R}. 
The number density of electrons integrated over the interval 3.857
to 8.5~eV is   9.9$\times10^{-8}$~cm$^{-3}$. 
We  find a production rate for NH$_2^-$ by reaction (\ref{NH2m from NH3}) of 
\begin{equation}
\int n_\mathrm{e}(\varepsilon)\,\sigma_\mathrm{DA}(\varepsilon)\,\upsilon\,\mathrm{d}\varepsilon =  7.0\times10^{-18}\ \ \mathrm{[s^{-1}\ per\ NH_3}]\ . 
\end{equation}
 
The possible destruction processes of NH$_2^-$ include photo-detachment, reactions with H$_2$, and 
mutual neutralisation in reactions with the most abundant
positive ions. 
The reaction with H$_2$, 
 \begin{equation}\label{NH2m from NH3}
\mathrm{NH_2^-} + \mathrm{H}_2 \rightarrow \mathrm{NH}_3 + \mathrm{H^-} \ , 
\end{equation}
is known to be rapid at low temperatures around 20~K but to decrease with
increasing temperature \citep{2008PhRvL.101f3201O}. 
If the reaction with H$_2$ is the dominant loss
process,  then the density of  NH$_2^-$ at $T=50-100$~K will be of the order of
\begin{equation}
n(\mathrm{NH}_2^-) \sim 7\times 10^{-8}\,\frac{n(\mathrm{NH}_3)}{n(\mathrm{H}_2)}\ \ \mathrm{[cm^{-3}]}\ , 
\end{equation}
which immediately suggests a very low NH$_2^-$ abundance. 

A second formation route  of NH$_{2}^{-}$ is via slow radiative attachment of electrons to NH$_{2}$  
\begin{equation}  
{\rm NH_{2} + e^{-} \longrightarrow NH_{2}^{-}   + h\nu}\ . 
\end{equation}
   Radiative attachment via emission from excited vibrational states has been calculated to be an efficient 
process to produce negative molecular ions with large electron affinities \mbox{(3-4 eV)} and at least four atoms
 \citep{2008ApJ...679.1670H}.   The NH$_{2}^{-}$ anion is smaller than this limit and has only a moderate 
electron affinity of 0.771 eV 
\citep{1989JChPh..91.2762W}.  Using Eq.~(11) in \citet{2008ApJ...679.1670H}, we estimate the rate coefficient 
for radiative attachment via the vibrational mechanism to be  only 
\mbox{$1 \times 10^{-17} (T/ 300 {\rm K})^{-1/2}$~cm$^{3}$~s$^{-1}$}.    The process is at most competitive with 
dissociative attachment of NH$_3$ via non-thermal electrons, and does not change the conclusion that the 
NH$_{2}^{-}$ anion has a low abundance.  Finally, if the anion is formed in a local region rich in atomic 
rather than molecular hydrogen, it can be destroyed by associative attachment with atomic hydrogen, 
 \begin{equation}
  {\rm NH_{2}^{-} + H   \longrightarrow NH_{3} + e^{-},}
  \end{equation}
   or competitively via photo-detachment.  

\section{Chemical models}

 \begin{figure} 
\resizebox{\hsize}{!}{ 
\includegraphics[angle=-90]{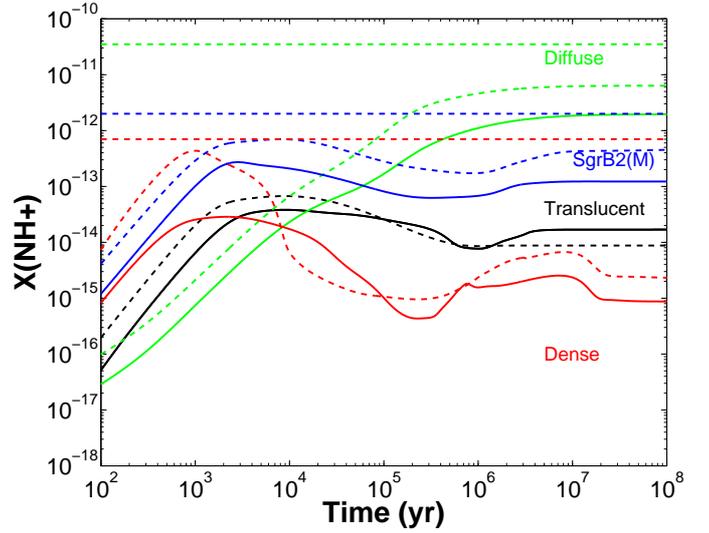}}
\caption{Temporal evolution of the NH$^+$ abundance for all four models 
(Table~\ref{Table: chemical model parameters}) where each model is plotted with   a 
\emph{high metal abundance}  (dashed lines) and  with a \emph{low metal abundance}  (solid lines). 
The translucent and dense models are plotted for  
\mbox{$T_\mathrm{K} = 30$~K} alone.  The \emph{observed}   upper limits   are indicated with   dashed horizontal 
lines  following the respective model colour code.}
\label{onlineFigB7}
\end{figure}

\begin{figure*}[b]
\centering
\subfigure{
\includegraphics[angle=-90,scale=.35]{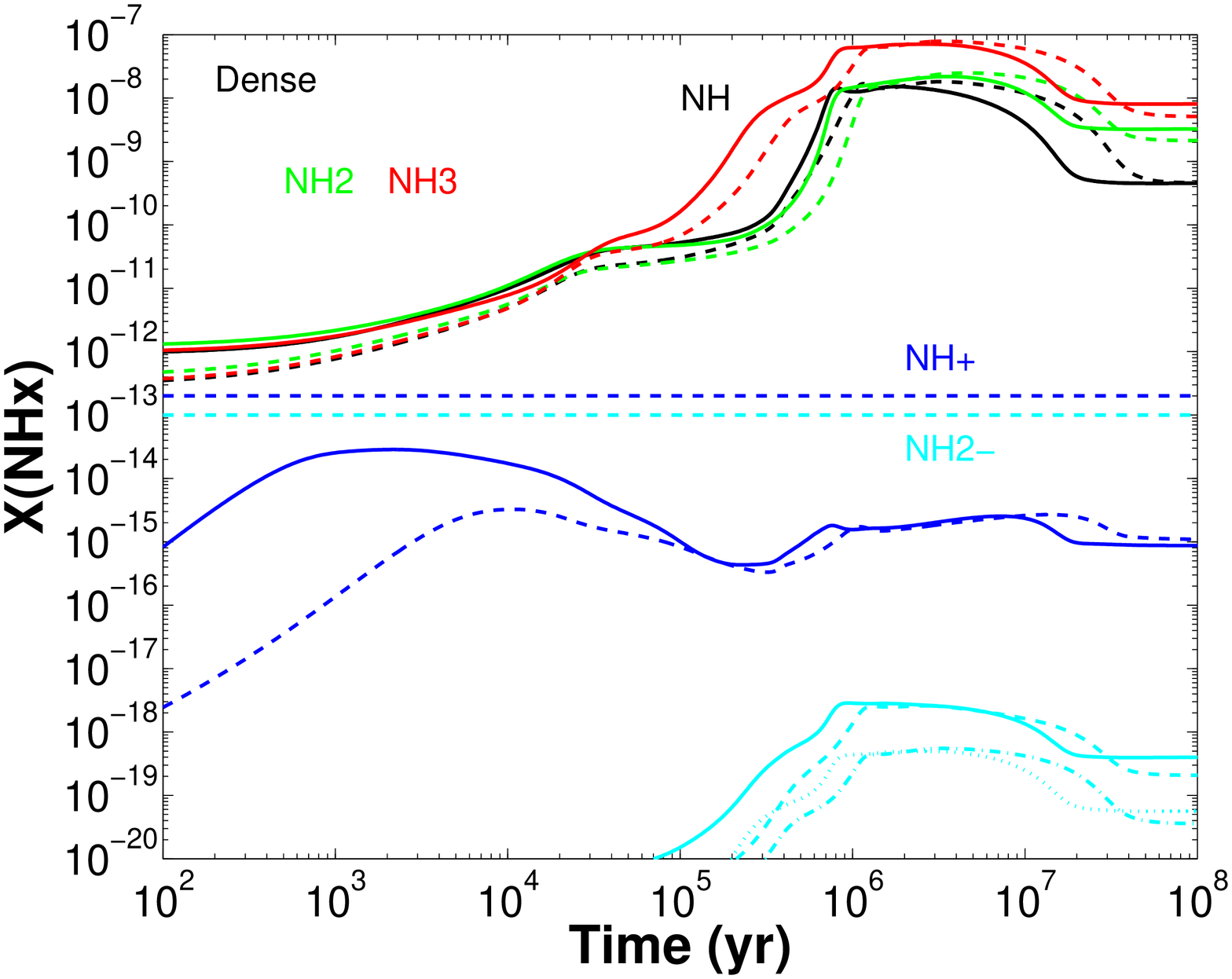}}
\subfigure{
\includegraphics[angle=-90,scale=.35]{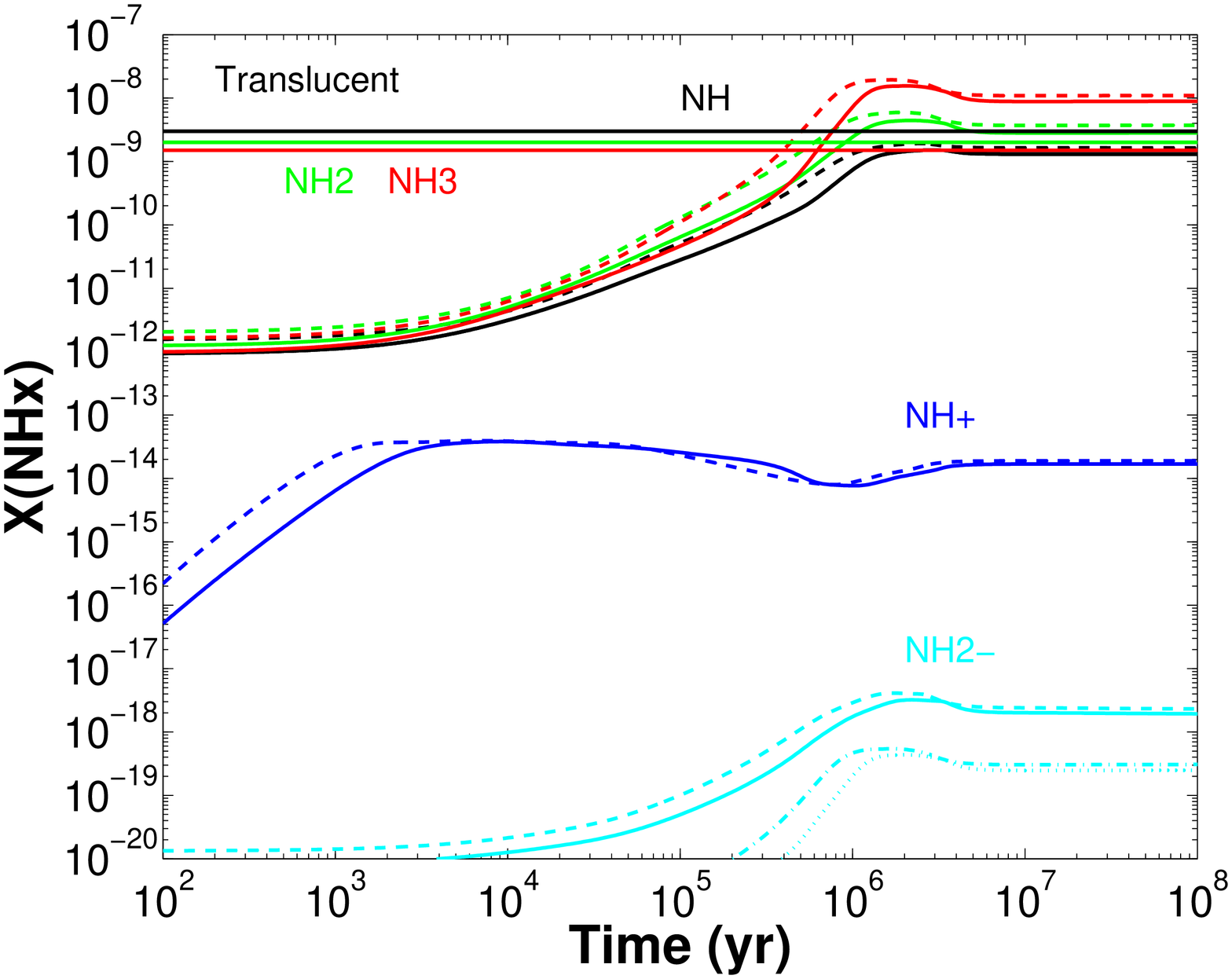}}
\vspace{.3in}
\subfigure{
\includegraphics[angle=-90,scale=.35]{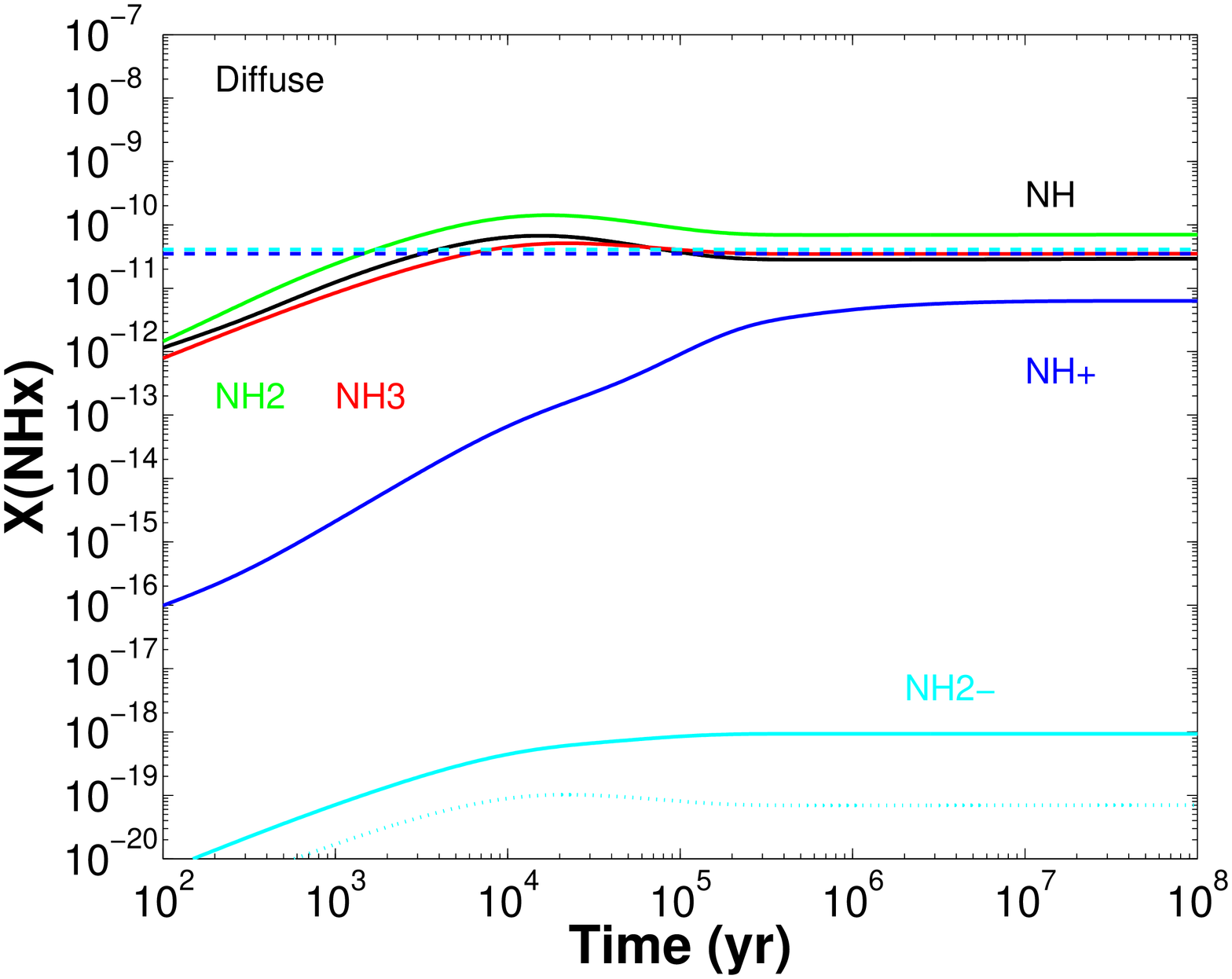}}
\subfigure{
\includegraphics[angle=-90,scale=.35]{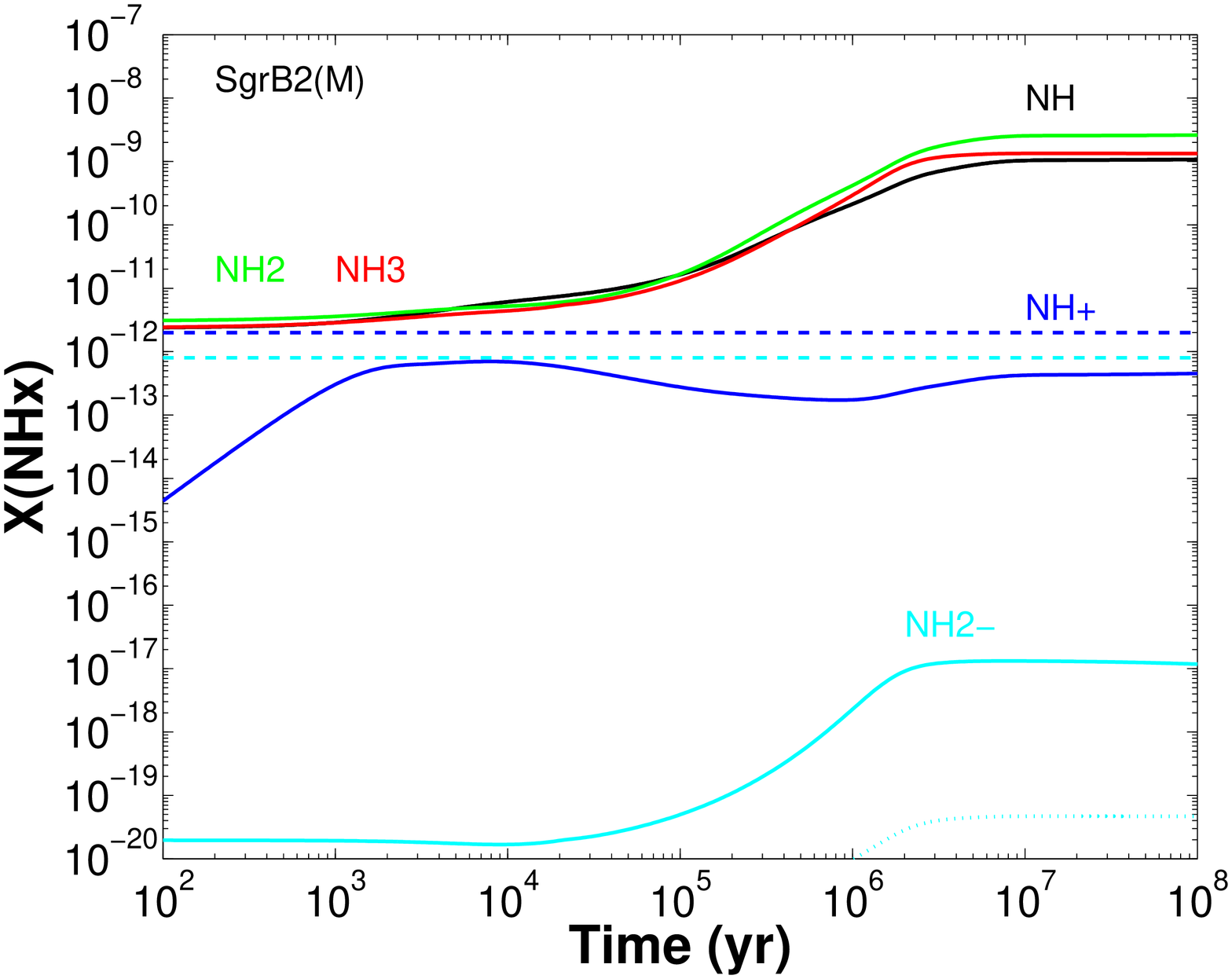}}
\caption{Temporal evolution of the nitrogen hydride abundances. 
Upper left:  
\emph{dense gas}; 
Upper right:  \emph{translucent gas};  
 Lower left: \emph{diffuse gas}; 
Lower right: \emph{Sgr\,B2\,(M) envelope model}   (see Table~\ref{Table: chemical model parameters}). 
The dot-dashed and dotted lines for NH$_2^-$  represent the 
estimated $X$(NH$_2^-)\sim7\times 10^{-8}\times X(\mathrm{NH}_3)$   for the 
$T_\mathrm{K} = 30$~K and 50~K (translucent) and $T_\mathrm{K} = 30$~K and 10~K (dense) models.
The \emph{observed} abundances  and upper limits corresponding to the different models are indicated with solid and dashed horizontal 
lines, respectively, following the respective species colour code.  
}
\label{onlineFigB4}
\end{figure*}
 
\begin{figure*}[b]
\centering
\subfigure{
\includegraphics[angle=-90,scale=.34]{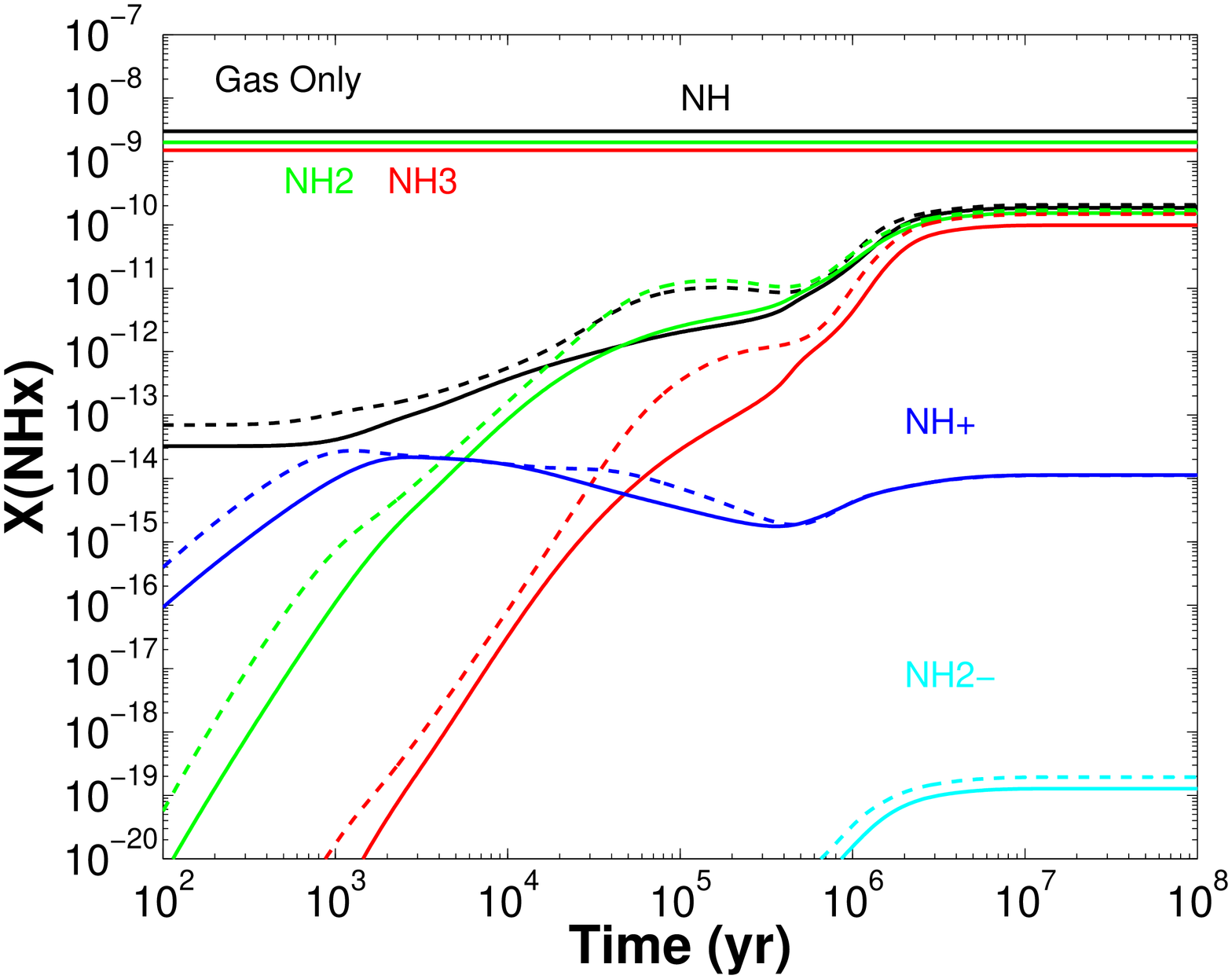}} 
\hspace{.3in}
\subfigure{
\includegraphics[angle=-90,scale=.34]{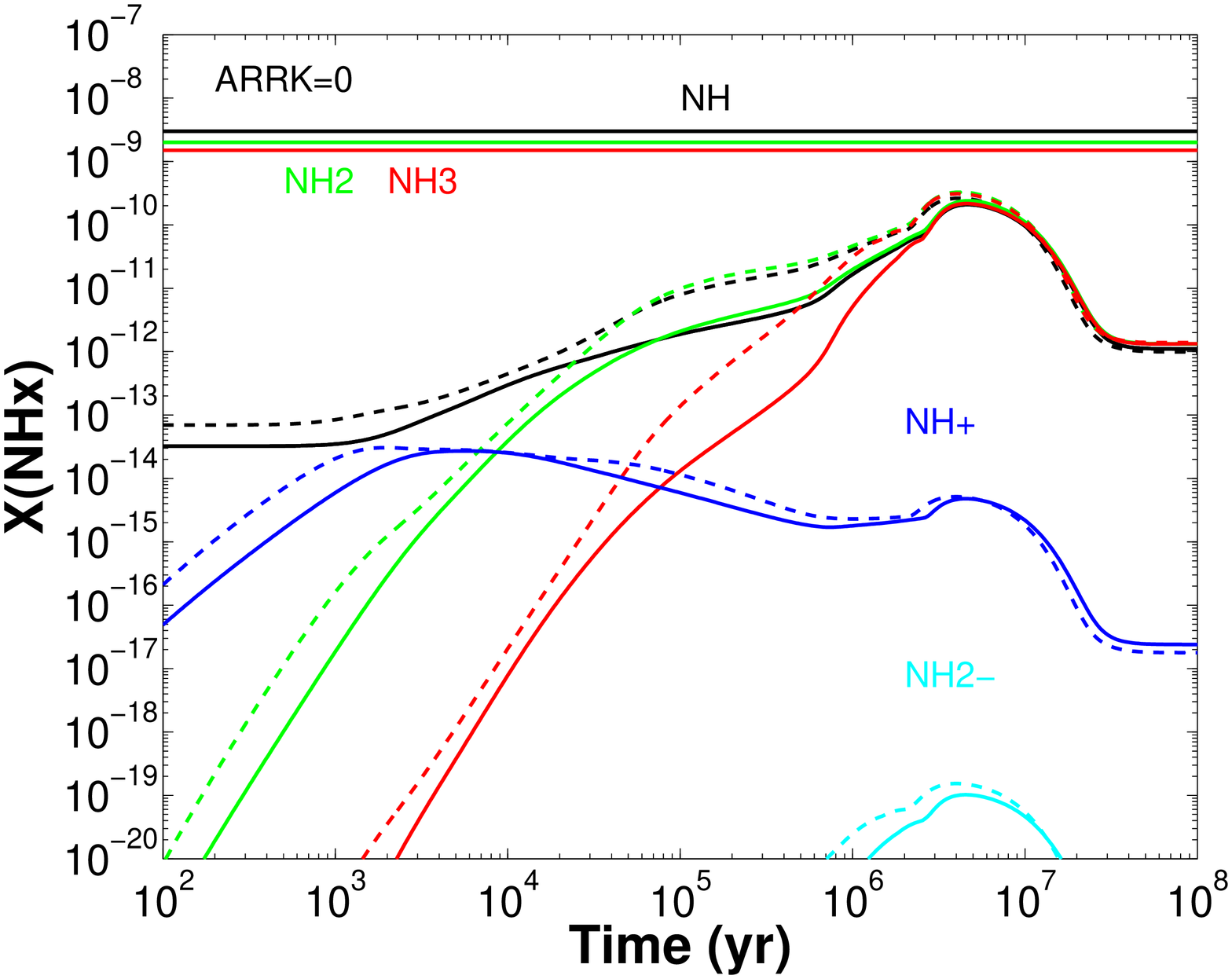}}   
\vspace{.3in}
\subfigure{
\includegraphics[angle=-90,scale=.34]{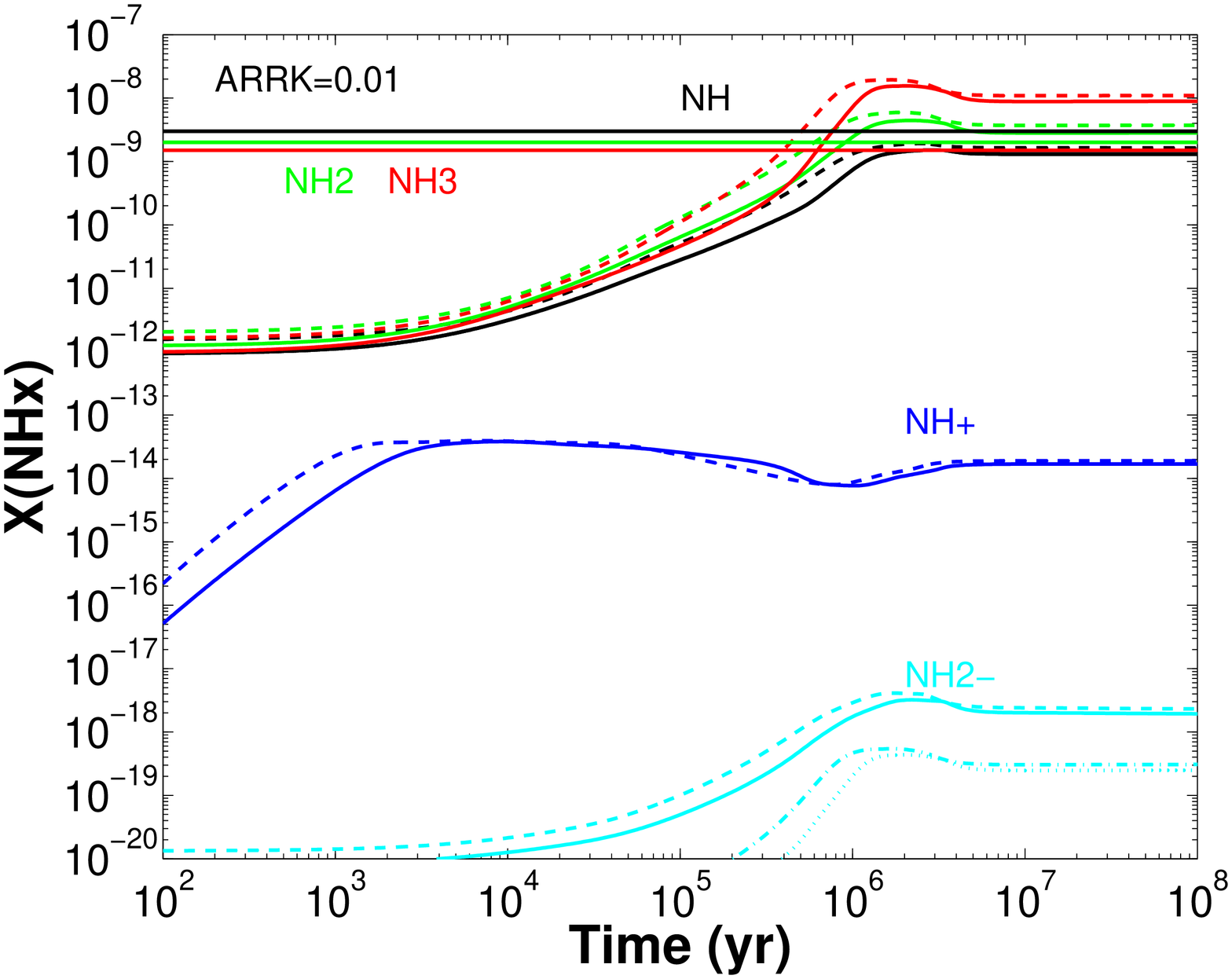}} 
 \hspace{.3in}
\subfigure{
\includegraphics[angle=-90,scale=.34]{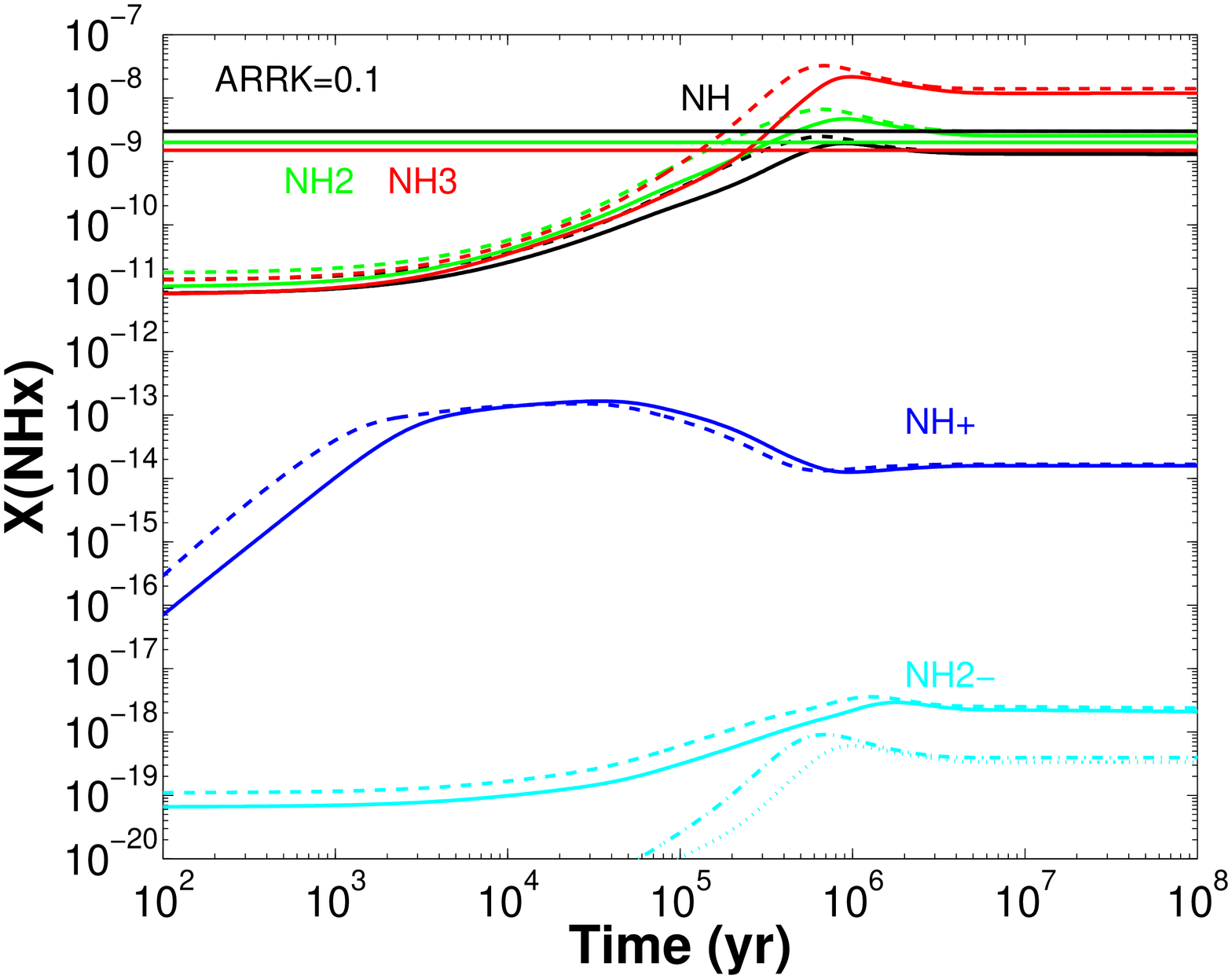}}
\caption{\emph{Translucent gas conditions} in all models (see Table~\ref{Table: chemical model parameters}).  
Upper left: Pure gas phase chemistry.
Upper right: Gas and surface chemistry and 
inactive non-thermal desorption efficiency \mbox{$a_\mathrm{RRK} = 0$}.
Lower left:   Gas and surface chemistry 
and active 
non-thermal desorption with the typical efficiency \mbox{$ a_\mathrm{RRK} = 0.01$}.  
Lower right: Gas and surface chemistry 
and high active 
non-thermal desorption (\mbox{$a_\mathrm{RRK} = 0.1$}).
  The \emph{observed} abundances and upper limits are indicated with solid and dashed horizontal 
lines, respectively, following the respective species colour code. 
The dot-dashed and dotted lines for NH$_2^-$  represent the 
estimated $X$(NH$_2^-)\sim7\times 10^{-8}\times X(\mathrm{NH}_3)$   for the 
$T_\mathrm{K} = 30$~K and 50~K models. 
If the reactive desorption mechanism is active with the typical 
$a_\mathrm{RRK} = 0.01$, each NH, NH$_2$  and NH$_3$ species that is formed on the grain through a
 hydrogenation reaction has a probability of $9.3 \times 10^{-3}$, $7.6 \times 10^{-3}$, and 
$5.2 \times 10^{-3}$, respectively, to desorb into the gas phase. There it will become available for 
detection and for follow-up reactions. Experiments by \citet{Dulieu:2013} indicate that this type 
of non-thermal desorption could be much more efficient on bare grains than  $a_\mathrm{RRK} = 0.01$. 
As shown, our models are not very sensitive to the exact value of the desorption probability, since the  model  
 with  $a_\mathrm{RRK} = 0.1$  gives very similar 
results to the model using $a_\mathrm{RRK} = 0.01$.
}
\label{onlineFigB5}
\end{figure*}
\clearpage

 \begin{figure} 
\resizebox{\hsize}{!}{ 
\includegraphics[angle=-90]{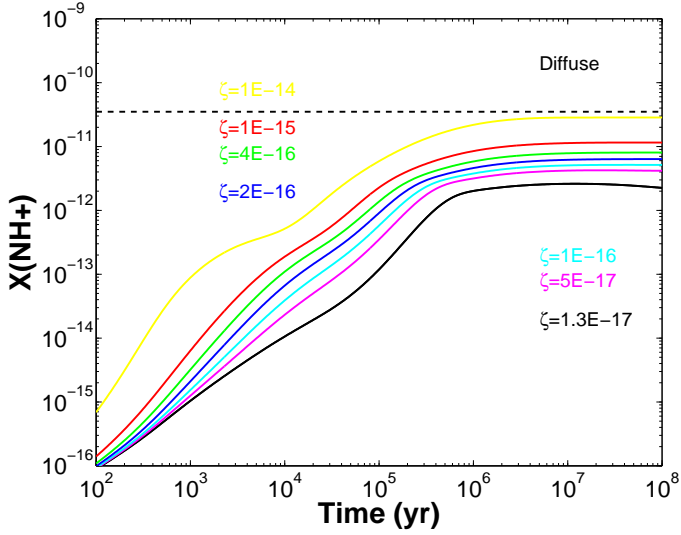}}  % diffuse high metal initial conditions
\caption{Temporal evolution of the NH$^+$ abundance for different 
cosmic ray ionisation rates for \emph{diffuse gas conditions} (Table~\ref{Table: chemical model parameters}). The dashed black line represents the observed upper NH$^+$ limit in
the diffuse gas.} 
\label{onlineFigB6}
\end{figure}

\section{Tables}

\begin{table}[\!htb] 
\centering
\caption{Hyperfine structure components of NH$^+$   \mbox{$^2\Pi_{1/2}$ $N = 1 - 1, J = 1.5^- - 0.5^+$}. 
}
\begin{tabular} {cccc c } 
 \hline\hline
     \noalign{\smallskip}

 Frequency	   & $A_{ul}$& g$_u$  & $\Delta v $\tablefootmark{a}    & Rel. Intensity\tablefootmark{b}  \\    \noalign{\smallskip}  %&  Band 
(GHz) &(s$^{-1}$)&  &(km\,s$^{-1}$) & $\frac {A_{ul}\times g_u} {A_{ul}\mathrm{(main)}\times g_u \mathrm{(main)}}$  \\
     \noalign{\smallskip}
     \hline
     \noalign{\smallskip}
 
 1\,012.516    &  0.00885  & 5 &  1.9 &   0.47 \\
 1\,012.523    &  0.01358   & 7 &0  & 1\\
 1\,012.529    & 0.00903    & 3 &  -1.8  &  0.28  \\
 1\,012.533    & 0.00193    & 3 &  -3.0 &  0.06   \\
 1\,012.534     & 0.00208   &5 &  -3.4  &  0.11 \\
 1\,012.550     & 0.00265    &5  &  -8.1  &  0.14  \\
 1\,012.556    & 0.00471    &5  &  -10 &  0.25 \\
 1\,012.556    & 0.00703    &3 &   -10  &  0.22   \\
 1\,012.567    & 0.00261    &3 &   -13  &  0.08  \\
 1\,012.570    & 0.00887    & 1&  -14  &  0.09   \\
 1\,012.571     & 0.00857   & 5&  -14   &  0.45   \\
 1\,012.574    & 0.00226    & 3 &   -15  &  0.07   \\
 1\,012.589   & 0.00427  & 3 &  -20 &  0.13  \\
 1\,012.604    & 0.00471    & 1 &   -24  &  0.05   \\

     \noalign{\smallskip}
\hline 
\label{Table: 1013 hfs transitions}
\end{tabular}
\tablefoot{ 
\tablefoottext{a}{The velocity offset from the strongest hfs component at 1\,012.523~GHz.} 
\tablefoottext{b}{The sum of the relative intensities  
of the 14 hfs components is 3.4.}  
}
\end{table}

\begin{table}[\!htb] 
\centering
\caption{Hyperfine structure components of NH$^+$   \mbox{$^2\Pi_{1/2}$ $N = 1 - 1, J = 1.5^+ - 0.5^-$}   
}
\begin{tabular} {cccc c } 
 \hline\hline
     \noalign{\smallskip}

 Frequency	   & $A_{ul}$& g$_u$  & $\Delta v $\tablefootmark{a}    & Rel. Intensity\tablefootmark{b}  \\    \noalign{\smallskip}  %&  Band 
(GHz) &(s$^{-1}$)&  &(km\,s$^{-1}$) & $\frac {A_{ul}\times g_u} {A_{ul}\mathrm{(main)}\times g_u \mathrm{(main)}}$  \\
     \noalign{\smallskip}
     \hline
     \noalign{\smallskip}
 
  1\,018.911 &   0.00137 &  3 &   83  & 0.04 \\
 1\,019.013 &  0.00095 &3  & 53   & 0.03 \\
 1\,019.020 &  0.00566 &5 &  51  &   0.29  \\
 1\,019.060 &  0.00434 &  3 &  39  & 0.13  \\
 1\,019.067 &  0.00113 &5  & 37   & 0.06 \\
 1\,019.184 &  0.00130 &5  &  2.5   & 0.07 \\
 1\,019.193 &  0.01392 &7  &               0      & 1.00 \\ % strongest
 1\,019.229 &  0.00381 & 3 &  -11  & 0.12 \\
 1\,019.232 &  0.01247 &5 &   -11  &   0.64 \\
 1\,019.251 &  0.01328 &1 &   -17  &   0.14 \\
 1\,019.259 &  0.00813 & 3&    -20&    0.25  \\
 1\,019.330 &  0.00547 & 3 &   -40 &    0.17 \\
 1\,019.361 &  0.00315 &  3 &    -49 &  0.10 \\
 1\,019.368 &  0.00713 &   5&    -51 &   0.37  \\

       \noalign{\smallskip}
\hline 
\label{Table: 1019 hfs transitions}
\end{tabular}
\tablefoot{ 
\tablefoottext{a}{The velocity offset from the strongest hfs component at 1\,019.193~GHz.} 
\tablefoottext{b}{The sum of the relative intensities  
of the 14 hfs components is 3.4.}  
}
\end{table}

\begin{table}[\!htb] 
\centering
\caption{Para-NH$_2^-$  $^1A_1$, \mbox{$J_{K_a,K_c} = 1_{1,1} - 0_{0,0}$}. 
Spectroscopic data from \citet{1986JChPh..85.4222T}. See Sect.~\ref{Section: observations and data reduction} for more details. 
}
\begin{tabular} {cccc} 
 \hline\hline
     \noalign{\smallskip}

Frequency\tablefootmark{a} &  error\tablefootmark{b}	    &  $A_{ul}$\tablefootmark{c}& $E_\mathrm{u}$ 	  \\    \noalign{\smallskip}   
(GHz)  &(MHz) &(s$^{-1}$)& (K)  \\
     \noalign{\smallskip}
     \hline
     \noalign{\smallskip}
 
933.855& $\gtrsim100$ & 5.43e-03 & 45  \\

     \noalign{\smallskip}
\hline 
\label{Table: NH2m hfs transitions}
\end{tabular}
\tablefoot{ 
\tablefoottext{a}{\citet{2011.ECLA.cernicharo}.}
\tablefoottext{b}{Error of predicted frequency.}  
\tablefoottext{b}{Using a ground state dipole moment of 1.311~Debye \citep[estimated uncertainty is 0.01~D;][]{1993JChPh..99.8349B}.}
}
\end{table} 

\clearpage
 
 \begin{table*}[\!htb] 
\centering
\caption{\emph{Herschel} OBSID's of the observed transitions analysed in this paper.
}
\begin{tabular} {llrccccc } 
 \hline\hline
     \noalign{\smallskip}
Source & Species 	& Frequency & Band & LO-setting$^a$ & Date &	OBSID    \\    \noalign{\smallskip}

&  & (GHz)    \\
     \noalign{\smallskip}
     \hline
\noalign{\smallskip}

\noalign{\smallskip} \noalign{\smallskip} \noalign{\smallskip} 
%-----------------------------------------------------------------
G10.6-0.4 & NH$^+$     & 1\,012.540 & 4a & A & 2012-04-10  & 1342244052  \\ 
	  &            &            &    & B &             & 1342244053   \\
	  &            &            &    & C &             & 1342244054   \\

\noalign{\smallskip} 
     &            & 1\,019.210 & 4a & A & 2012-04-10  &  1342244055 \\  
     &            &            &    & B &             &  1342244056  \\
     &            &            &    & C &             &  1342244057  \\

\noalign{\smallskip}

     & p-NH$_2^-$ & 933.855 & 3b  & A &2012-09-18  &  1342251113 \\  
     &          &         &     & B &            &  1342251114 \ \\
     &          &         &     & C &            &  1342251115 \ \\

\noalign{\smallskip} \noalign{\smallskip} \noalign{\smallskip} 
%-----------------------------------------------------------------

%-----------------------------------------------------------------
Sgr\,B2\,(M) & NH$^+$     & 1\,012.540 & 4a & SScan & 2012-04-04  &  1342243701 \\

\noalign{\smallskip} 
 &      & 1\,019.210 & 4a & SScan & 2012-04-04  & 1342243702 \\  
 
\noalign{\smallskip}

     & p-NH$_2^-$ & 933.855 & 3b  & SScan &2012-09-18  &  1342251112 \\

\noalign{\smallskip} \noalign{\smallskip} \noalign{\smallskip} 
%-----------------------------------------------------------------

    \noalign{\smallskip}
\hline 
\label{Table: obsid}
\end{tabular}
\tablefoot{
\tablefoottext{a}{Three different frequency settings of the LO were performed towards G10.6$-$0.4,  
with approximately 15~km~s$^{-1}$ 
between each setting in order to determine the  sideband origin of the signals. Towards Sgr\,B2\,(M) we used 
the spectral scan mode with 8 different LO settings.}  
}
\end{table*} 

\clearpage

\begin{table}
\caption{Initial elemental abundances}
\center
\begin{tabular}{lcc}
\hline\hline
 \noalign{\smallskip}
Species & $ n_{i}/n_{\rm H}$ & $ n_{i}/n_{\rm H}$\\ 
$i$ & high metal$^{a,b}$ & low metal$^{a,c}$ \\
     \noalign{\smallskip}
     \hline
\noalign{\smallskip} 
H & 1 & $\dots$\\
H$_2$ & $\dots$ & 0.5\\
He  & 0.09& 0.14\\
C$^+$ & 1.4(-4)& 7.3(-5) \\
N & 7.5(-5)& 2.14(-5) \\
O & 3.2(-4)& 1.76(-4) \\
S$^+$ & 1.5(-6)& 8.0(-8) \\
Na$^+$ & 2.0(-8)& 2.0(-9) \\
Mg$^+$ & 2.55(-6)& 7.0(-9) \\
Si$^+$ & 1.95(-6)& 8.0(-9) \\
P$^+$ & 2.3(-8)& 3.0(-9) \\
Cl$^+$ & 1.4(-8)& 4.0(-9) \\
Fe$^+$ & 7.4(-7)& 3.0(-9) \\
\hline
\end{tabular}
\label{Initial elemental abundances}
\tablefoot{
\tablefoottext{a}{$x(y)=x\times10^y$} 
\tablefoottext{b}{Adopted from \citet{2006A&A...457..927G} and \citet{2008ApJ...680..371W}}.
\tablefoottext{c}{Adopted from \citet{2007A&A...467.1103G}, based on 
\citet{1982ApJS...48..321G}.}
}
\end{table}

\end{document}